\pdfoutput=1
\documentclass[lettersize,journal]{IEEEtran}
\usepackage{amsmath,amsfonts}
\usepackage{algorithmic}
\usepackage{algorithm}
\usepackage{array}
\usepackage[caption=false,font=scriptsize,labelfont=rm,textfont=rm]{subfig}
\usepackage{textcomp}
\usepackage{stfloats}
\usepackage{url}
\usepackage{verbatim}
\usepackage{graphicx}
\usepackage{cite}
\usepackage{bm}
\usepackage{pifont}
\usepackage{makecell}
\usepackage{bigstrut}
\usepackage{multirow}
\usepackage{makecell}
\usepackage{color}
\makeatletter

\newcommand{\Rmnum}[1]{\expandafter\@slowromancap\romannumeral #1@}
\makeatother

\begin{document}

\title{Compression before Fusion: Broadcast Semantic Communication System for Heterogeneous Tasks}

\author{Mingze Gong,~\IEEEmembership{Graduate Student Member,~IEEE}, Shuoyao Wang,~\IEEEmembership{Member,~IEEE}, Fangwei Ye,~\IEEEmembership{Member,~IEEE}, \\
        and Suzhi Bi,~\IEEEmembership{Senior Member,~IEEE}
\thanks{Part of this work \cite{gong2023scalable} has been presented in IEEE Global Communications Conference 2023. }
}

\maketitle

\begin{abstract}
        Semantic communication has emerged as new paradigm shifts in 6G from the conventional syntax-oriented communications. 
        Recently, the wireless broadcast technology has been introduced to support semantic communication system toward higher communication efficiency. 
        Nevertheless, existing broadcast semantic communication systems target on general representation within one stage and fail to balance the inference accuracy among users. 
        In this paper, the broadcast encoding process is decomposed into compression and fusion to improves communication efficiency with adaptation to tasks and channels.
        Particularly, we propose multiple task-channel-aware sub-encoders (TCE) and a channel-aware feature fusion sub-encoder (CFE) towards compression and fusion, respectively. 
        In TCEs, multiple local-channel-aware attention blocks are employed to extract and compress task-relevant information for each user. 
        In GFE, we introduce a global-channel-aware fine-tuning block to merge these compressed task-relevant signals into a compact broadcast signal. 
        Notably, we retrieve the bottleneck in DeepBroadcast and leverage information bottleneck theory to further optimize the parameter tuning of TCEs and CFE.
        We substantiate our approach through experiments on a range of heterogeneous tasks across various channels with additive white Gaussian noise (AWGN) channel, Rayleigh fading channel, and Rician fading channel.
        Simulation results evidence that the proposed DeepBroadcast outperforms the state-of-the-art methods. 
\end{abstract}


\section{Introduction}\label{introduction}
    \IEEEPARstart{s}{ixth} generation (6G) has been conceptualized as an intelligent information system both being driving and driven by the modern artificial intelligence (AI) technology\cite{letaief2019roadmap}. 
    Bridging the AI applications and physical world, semantic communication achieves reliable transmission with impressive semantic level information fidelity \cite{gunduz2022beyond}\cite{zhang2022toward} by extracting the latent semantic features from the source. 
    In fact, semantic communication has been recognized as promising technique to improve communication efficiency and breakthrough beyond Shannon paradigm \cite{qin2021semantic}.

    Recent years have witnessed the development of neural-network-based semantic communication systems, most of whom shared the core framework called joint source and channel coding (JSCC).  
    In particular, JSCC has been investigated as a solution to address the ``cliff effect'' \cite{niu2023hybrid} in separate source and channel coding (SSCC), and garnered growing interest in handling massive volume of data. 
    \begin{figure}[!t]
        \centering
        \includegraphics[scale=0.55]{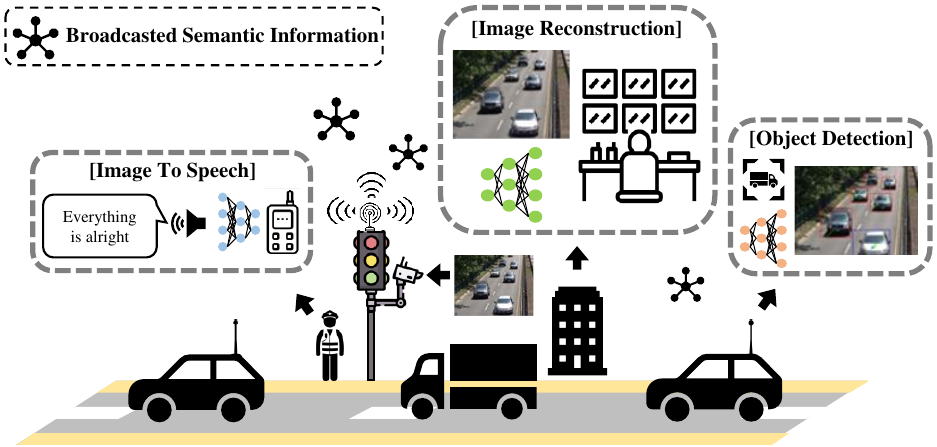}
        \caption{A semantic communication system with single server and multiple devices. }
        \label{fig:practicalscenario}
    \end{figure}
    In most recently, one-to-one systems i.e., one transmitter and one receiver, have successfully achieved higher communication efficiency \cite{bourtsoulatze2019deep, xie2021deep, xu2021wireless, xu2022deep} compared with the conventional efficient source coding (e.g., BPG, JPEG, JPEG2000) and near-optimal channel coding (e.g., LDPC, Turbo Codes, Polar Codes). 
    
    Alternatively, the high scalability and communication efficiency of conventional wireless broadcast systems make wireless broadcast technology well-suited for handling the same service request from multiple users \cite{zheng2004spatial}. 
    Recently, broadcast technology has been extended to semantic communication and named \emph{broadcast}, also referred to \emph{one-to-many}, \emph{semantic communication system}.
    Broadcast semantic communication systems aim at fulfilling the requirements from multiple edge devices requesting for source information at the central server, which is shown in Fig.\ref{fig:practicalscenario}. 
    To guarantee transmission quality, it is crucial for the encoder to extract semantic information for multiple users and preserve it from physical noise during transmission. 
    However, as concept-of-proof explorations, most of the existing research has only considered simple scenarios where multiple users perform the same task in a single-type channel environment to study broadcast semantic communication.
    In practice, both the users' channels and the required tasks are likely to be heterogeneous.
    Under the complex conditions of heterogeneous channels and tasks, current approaches face significant performance degradation.

    The exploration for supporting broadcast semantic communication system to adapt to heterogeneous and complex physical channels to meet requirements from distributed devices is indeed needed. 
    In this paper, we introduce DeepBroadcast, a multi-user broadcast task-oriented semantic communication system. 
    Differing from one-to-one semantic communication, in a broadcast system, the channel encoder must simultaneously accomplish the compression and encoding of task-relevant information for each task with balance to handle heterogeneous noise distortion. 
    The current channel encoder architecture struggles to meet these requirements.
    Consequently, we have deconstructed the channel encoder into compression and fusion and designed task-channel-aware sub-encoders (TCEs) and channel-aware feature fusion sub-encoder (CFE) to fulfill the aforementioned requirements.
    The major contributions are summarized as follows:
    \begin{enumerate}
    \item {} We propose a task-channel-aware broadcast semantic communication system for heterogeneous tasks over heterogeneous broadcast channels. We leverage information bottleneck (IB) theory and formulate the location of bottleneck in the broadcast system for system enhancement.
    \item {\emph{Decomposed Channel Encoder}}: Given the inherent variability in different tasks and channels, we deconstruct channel encoder into TCEs and CFE to jointly improves the communication efficiency with adaptation to tasks and channels. TCEs aims at autonomously extracting and condensing task-specific information into noise-robust signal independently for each user. Moreover, CFE empowers the transmitter to achieve better fusion of task-specific signals with less information redundancy among different tasks while ensure the distinct feature.
    \item {\emph{Superior Performance} }: Compared with the state-of-the-art task-oriented broadcast semantic communication scheme, DeepBroadcast achieves remarkable noise robustness and multi-task performance. Visualization results reveal the source of DeepBroadcast's superiority, shortening semantic distance between transmitted and received signals. Furthermore, simulation results respectively tested in two and three-user scenarios with heterogeneous channels and tasks demonstrate the outstanding performance of DeepBroadcast. 
    \end{enumerate}

    The remainder of this paper is organized as follows: 
    In the following Section~\ref{relatedworks}, we will introduce the recent related works and show the details of the proposed broadcast semantic communication system and the formulated problem in Section~\ref{systemmodel}.
    Section~\ref{proposedmethod} illustrates the architecture of the proposed system and describes the training algorithm.
    Then, we present the simulation results in Section~\ref{experiment}.
    Finally, Section~\ref{conclusion} summarizes this paper. 
    Notably, Table.~\ref{tab:notation} summarizes the notations in this paper. 
    
\begin{table}[!t]
    \centering
    \caption{The Summary of Notations}
    \resizebox{\linewidth}{!}{
      \begin{tabular}{ccccc}
      \hline
      Notation & \multicolumn{4}{c}{Definition} \bigstrut\\
      \hline
      $\bm{x}$     & \multicolumn{4}{c}{input of the system} \bigstrut[t]\\
      $\bm{y}_i$    & \multicolumn{4}{c}{label of $i$-th task} \\
      $\hat{\bm{y}}_i$ & \multicolumn{4}{c}{inference result in $i$-th receiver} \\
      $\phi, \theta$ & \multicolumn{4}{c}{trainable parameters} \\ 
      $\bm{z}^{e}_{i}$ & \multicolumn{4}{c}{\makecell{extracted semantic feature }} \\
      $\bm{z}^{r}_{i}$ & \multicolumn{4}{c}{\makecell{refined semantic feature}} \\
      $\bm{z}^{f}_{i}$ & \multicolumn{4}{c}{\makecell{fine-tuned semantic feature}} \\
      $SNR_i, r_i$ & \multicolumn{4}{c}{signal-noise ratio of $i$-th broadcast channel } \\
      $\overrightarrow{\bm{SNR}}$ & \multicolumn{4}{c}{signal-noise ratios of all broadcast channels } \\
      $\bm{m}^{\epsilon}, \bm{m}^{\rho}, \bm{m}^{g}$ & \multicolumn{4}{c}{modulated features} \\
      $\bm{z}$     & \multicolumn{4}{c}{broadcast signal} \\
      $\bm{h}_i$    & \multicolumn{4}{c}{channel gain of $i$-th broadcast channel} \\
      $\bm{n}_i$    & \multicolumn{4}{c}{probability distribution of $i$-th broadcast channel} \\
      $\hat{\bm{z}}_i$ & \multicolumn{4}{c}{received feature in i-th receiver} \\
      $\mathcal{L}$     & \multicolumn{4}{c}{loss function} \\
      $I(X,Y)$ & \multicolumn{4}{c}{mutual information between X and Y} \\
      $\mu, \sigma$    & \multicolumn{4}{c}{mean and standard deviation values of a Gaussian distribution} \\
      $p, q$  & \multicolumn{4}{c}{probability distributions} \\
      $\mathbb{E}(p)$     & \multicolumn{4}{c}{expectation of probability distribution $p$} \\
      $D_{KL}(p||q)$ & \multicolumn{4}{c}{Kullback-Leibler-divergence of distributions $p$ and $q$. } \\
      \hline
      \end{tabular}%
    }
    \label{tab:notation}%
  \end{table}%

\section{Related Works} \label{relatedworks}
    In this section, we briefly introduce recent works in semantic communication according to the number of users.
    The primary distinctions in application scenario between related works and our work are shown in Table.~\ref{tab:gaps}. 

\subsection{Semantic Communication}
    Semantic communication has been regarded as promising technique to achieve higher communication efficiency beyond Shannon paradigm \cite{shannon1948mathematical}. 
    Based on the system objectives, the existing works in semantic communication can be summarized into two categories. 
    Some of them aim at optimizing the system to minimize word/pixel level errors rather than bit errors, which can be named as \emph{reconstruction-oriented} semantic communication systems. 
    Besides, for solely task execution transmission between connected machines, some systems called \emph{task-oriented} semantic communication systems have also been developed to only transmit the task-related semantic information. 
    Alternatively, the number of users\footnote{In semantic communication, the term ``user'' can refer to either the transmitter, responsible for processing source information, or the receiver, who makes inference based on received semantic features. In this paper, we define ``user'' as the receiver, since the receiver is considered as the one truly utilizing semantic information for task execution or source data reconstruction. } is another crucial factor considered in designing both the conventional communication system as well as the semantic communication system. 

    Since 2019, semantic communications have achieved great success in \emph{one-to-one} systems \cite{bourtsoulatze2019deep, xie2021deep, xu2021wireless, xu2022deep}, i.e., from single transmitter to single receiver.    
    By jointly training the entire system with deep learning approaches, some reconstruction-oriented systems have achieved impressive noise robustness during transmission and performance in reconstructing the source data. 
    For instance, \cite{bourtsoulatze2019deep} developed a semantic communication system named DeepJSCC for image transmission, and \cite{xie2021deep} proposed a text transmission semantic communication system called DeepSC. 
    However, in some cases, not all receivers are expected to display source data such as an image or a line of text.
    Instead, some receivers are probably on duty of edge tasks.
    It is more efficient for them to directly infer based on received semantic features than the reconstructed source data. 
    Motivated by this, numerous task-oriented semantic communication systems \cite{zhang2022deep, hu2022robust, shao2021learning} have been proposed and demonstrated exceptional performance in specific tasks compared with the conventional SSCC approaches.
    Beyond individual task, some one-to-one systems \cite{sheng2022multi, zhang2022unified, ye2020deepnoma, lyu2024semantic} were explored to simultaneously infer for multiple tasks in the unified receiver. 

    To accommodate a growing number of distributed sensors, the conventional uplink multiple access is a widely adopted technology that allows multiple transmitters to multiplex the communication channel without causing interference.
    Recently, the uplink multiple access, which is also named \emph{many-to-one}, semantic communication system has raised significant attention \cite{lo2023collaborative, shao2022task, shao2022taskvideo}.
    By gathering multiple source information and handling them mainly in the individual receiver such as cloud server, the many-to-one semantic communication systems have successfully improved the utilization efficiency of communication resources. 
    
    Besides communications for single receiver, transmitting information to multiple receivers generally exists and the requirements are increasing in the real world. 
    For example, in the Internet of things, multiple devices are highly possible to simultaneously request for source information and execute tasks.
    Obviously, transmitting information to each receiver independently in such a network exhibits drawbacks in bandwidth usage and latency. 
    Recently, some researches such as \cite{li2023non} have explored the multi-user (i.e. multi-receiver) semantic communication systems, leveraging advantages of conventioanl wireless technologies (e.g., non-orthogonal multiple access and broadcast). 
    In particular, broadcast semantic communication systems \cite{ding2021snr, ma2023features, nguyen2023swin, wu2023fusion, sagduyu2023multi, sagduyu2023joint} have demonstrated impressive communication efficiency in transmitting single source information to multiple receivers. 

    \begin{table}[!t]
        \renewcommand{\arraystretch}{1.1}
        \centering
        \caption{The Application Scenarios of Related Works and Ours}
        \begin{tabular}{ccccc}
        \hline
        \multirow{3}[0]{*}{Ref} & \multicolumn{2}{c}{\multirow{2}{3cm}{\centering Heterogeneous Channel \\ Conditions}} & \multirow{3}{2cm}{\centering Heterogeneous \\ Tasks}\\
        \\
          & \multicolumn{1}{c}{\centering Channel Types} & \multicolumn{1}{c}{\centering Noise Intensity} \\
        \hline
        \cite{ding2021snr, ma2023features}      & \ding{55}     & \ding{51}  & \ding{55} \\
        \cite{nguyen2023swin}      & \ding{51}     & \ding{55}  & \ding{55} \\
        \cite{wu2023fusion, sagduyu2023multi, sagduyu2023joint}     & \ding{55}     & \ding{51}   & \ding{51} \\
        Ours     & \ding{51}     & \ding{51}  & \ding{51} \\
        \hline
        \end{tabular}%
        \label{tab:gaps}%
    \end{table}%

\subsection{Broadcast in Semantic Communication}    
    The high scalability and communication efficiency of conventional wireless broadcast systems make them suitable for accommodating service requests from multiple users \cite{zheng2004spatial}.
    Recently, this advantage has been extended to semantic communications. 
    However, challenges accompany the introduction of broadcast technology since broadcast channels are more complex.
    For instance, factors such as types and intensities of heterogeneous broadcast channels make the encoding much more challenging than that in single user communications. 
        
    Most recently, some broadcast semantic communication systems have been explored to address the problems. 
    Ref.~\cite{ding2021snr} proposed a multi-user JSCC system in broadcast manner for image transmission and firstly focuses on the adaptability for different signal-to-noise ratios (SNRs) in receiver side. 
    Considering intended semantic information among the users, \cite{ma2023features} proposed a broadcast semantic communication system to transmit certain semantic information to construct specific images. 
    Particularly, \cite{nguyen2023swin} indicated the variations of model architectures used by different users, which might influence the reconstruction quality of the source in degrees.
    Motivated by this, \cite{nguyen2023swin} has explored a reconstruction-oriented broadcast semantic communication system to transmit images to multiple receivers with decoding models in different capabilities.

    Unfortunately, previous works mainly focused on homogeneous task cases and could not be directly applied to the heterogeneous task cases.
    In order to transmit different images to two users in the efficient broadcast way, \cite{wu2023fusion} designed a fusion-based two-user semantic communication system for wireless image transmission over broadcast channels. 
    However, \cite{wu2023fusion} is only applicable in a specific system where there is one user with a good channel and another user with a poor channel. 
    For multiple users, a task-oriented multi-user semantic communication system called MTOC is proposed in \cite{sagduyu2023multi}.
    Impressively, under the setting of a transmitter and multiple receivers, MTOC leveraged the broadcast nature to simultaneously achieve multiple classification tasks through efficient communications. 
    Based on MTOC, \cite{sagduyu2023joint} further integrated semantic communications with joint sensing and communications in multi-task scenario. 

    Nevertheless, \emph{training the deep models originally designed for one-to-one semantic communication with broadcast channels suffers significant performance degradation when the broadcast channels are heterogeneous}.
    On the one hand, transmitter needs a complex encoding pattern to construct robust signal for heterogeneous channels.
    On the other hand, substantial and precise task-specific information should be simultaneously contained in the transmitted signal to accurately response different downstream tasks. 
    Thus, devices in charge of complex tasks exhibit lower tolerance to noise distortion, making heterogeneous broadcast channels even more threatening. 
    As shown in Section~\ref{experiment}, the broadcast semantic communication systems in \cite{sagduyu2023multi,sagduyu2023joint} fails to stabilize multi-task performance in heterogeneous broadcast channels. 
    As an important role in wireless communications, heterogeneous broadcast channels call for adapting semantic information to complex and variable noise distortions during transmission. 
    
    In this paper, \emph{we develop a task-oriented broadcast semantic communication system to effectively execute multiple tasks for multiple users over heterogeneous broadcast channels.}
    Particularly, we employ multiple task-channel-aware sub-encoders (TCEs) and a channel-aware feature fusion sub-encoder (CFE) in transmitter to collaboratively achieve the adaptation to heterogeneous channels.

\subsection{Channel-aware Encoder in Semantic Communication}
    Encoding the channel state information (CSI) into the semantic information space is a viable approach to preserve the transmitted signal from being corrupted by dynamic physical noise \cite{wu2023fusion,nguyen2023swin, xu2021wireless, xu2022deep}. 
    Besides fundamental JSCC architecture, the channel-aware encoders have been developed to improve the communication efficiency for semantic communications \cite{wu2023fusion, nguyen2023swin}.
    For instance, \cite{wu2023fusion} utilize the channel-aware encoder to adapt the length of transmitted signal with hard decision. 
    In ``hard decision'', we mean to adapt the signal length by masking unimportant feature. 
    However, the unimportant information should not be equated with the useless information. 
    Unlike hard decisions, soft decisions aim to emphasize the important semantic features without altering the code length. 
    And it is accomplished by adjusting the feature values based on side information. 
    Consequently, some semantic communication systems \cite{xu2022deep,xu2021wireless} have investigated soft decisions and incorporated them into the encoder. 
    
    In this paper, we adopt soft decisions for TCEs and CFE to adapt broadcast signal to heterogeneous broadcast channels.
    \emph{Besides following previous works to encode each CSI into corresponding task-relevant feature in particular TCE, we simultaneously combine all broadcast CSI and refined features within one module. }
    With joint processing of TCEs and CFE, the proposed system can effectively monitor the dynamic broadcast channels and attain commendable noise robustness.

\subsection{Information Bottleneck in Wireless Communication}
    The IB theory \cite{alemi2016deep} has provided a special insight of information extraction and compression. 
    The IB principle revealed that a good extracted feature should be efficient for the task inference while retaining minimal information from the source \cite{shao2021learning}.
    It is a trade-off between information extraction and information compression. 
    
    It is not easy to fuse different task-oriented information into one broadcast signal with limited information space. 
    As multiple TCEs are implemented to meet the customized requirements from tasks, it is efficient to transmit all task-specific information within single signal. 
    However, since semantic information for different tasks can be various in degrees, the information fusion becomes challenging within limited information space. 
    As shown in Section~\ref{experiment}, inappropriate objective function for training fails to balance different task-specific information and leads to multi-task performance degradation. 
    It is important to achieve good information compression among task-specific information before fusion.     
    
    Fortunately, the advantage of IB in information compression motivates us to meet the challenge with IB. 
    Additionally, the recent remarkable achievement of IB theory in semantic communications \cite{shao2022task, shao2022taskvideo, shao2021learning} encourages us to take a step forward with it. 
    However, \emph{most of existing works focus on leveraging IB to provide high-quality communications for single receiver.}
    Furthermore, it is challenging to directly employ the variational IB in broadcast semantic communication system, where transmitter communicates with multiple receivers simultaneously. 
    For example, where is the bottleneck within the broadcast system is still unclear.
    Therefore, a variational IB principle should be explored for broadcast semantic communication system. 
    In this paper, \emph{we extend the IB theory and apply it to the broadcast semantic communication system for multi-task execution and adapting to broadcast signal to heterogeneous channels. }

\section{Background and Motivation} \label{systemmodel}
\begin{figure}[!t]
    \centering
    \includegraphics[scale=0.35]{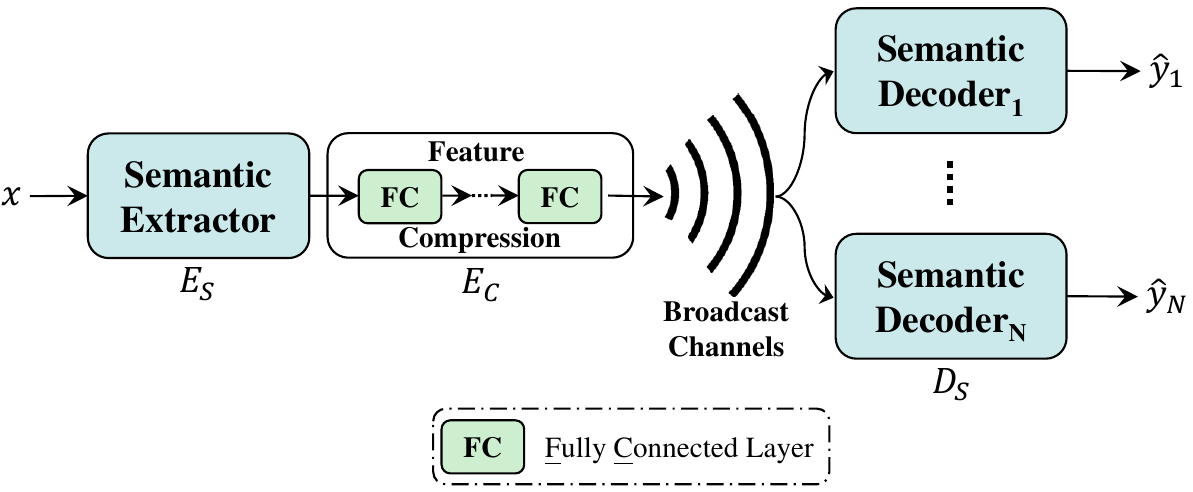}
    \caption{The general task-oriented broadcast semantic communication system model.}
    \label{fig:SEMD}
\end{figure}

	In this section, we will introduce the system model of task-oriented multi-user broadcast semantic communication system. 
    As shown in Fig.\ref{fig:SEMD}, we consider a multi-user broadcast semantic communication system for single source information transmission.
    The system consists of one server and $N$ edge devices, each of whom serve for one unique task respectively.
    Inspired by conventional broadcast communication systems, the main idea is to utilize JSCC method to generate the broadcast signal $Z$. 
    Without loss of generality, we denote that the inputs of the semantic communication system are images $\bm{x} \in \mathbb{R}^{c \times h \times w}$ with the number of color channels $c$ as well as the height $h$ and the width $w$ of the image signal, and its target $\bm{y}$.
    Here, $\bm{y} = \left[y_1, y_2, ..., y_N\right]$ and $y_i$ denotes the ground-truth label of $i$-th user. 
    The $\bm{x}$ and corresponding $\bm{y}$ are deemed as different realizations of a pair of random variables $(X,Y)$. 
    Then, the $i$-th user receives the signal $\hat{Z}_i$ and infers the task target $\hat{Y}_i$. 
    In the case of the $i$-th device, the inference result can be instantiated by random variables $\hat{Y}_i$. 
    In this way, the system can transmit semantic information and execute multiple tasks on different receivers simultaneously over a shared communication channel with the same frequency. 
    Notably, these random variables follow the Markov Chain:
        \begin{equation}
            \setlength{\arraycolsep}{1.2pt}
            Y \to X \to Z \to [\hat{Z}_1, \dots, \hat{Z}_i] \to [\hat{Y}_1, \dots, \hat{Y}_i]. \label{markovchain}
        \end{equation}
    Following the data stream, we will introduce the system model in the following. 

\subsection{System Model}

    Following \cite{sagduyu2023multi}, the system model contains one transmitter and multiple receivers. 
\subsubsection{Transmitter}
    Transmitter aims at extracting semantic information from source information to improve communication efficiency as much as possible. 
    As shown in Fig.~\ref{fig:SEMD}, the transmitter processes source information with a semantic extractor and multiple serial nerual networks (NNs). 
    In particular, transmitter extracts shared semantic information through the learned function $E_{S} :: \mathbb{R}^{c \times h \times w} \to \mathbb{R}^{c_{1} \times h_{1} \times w_{1}}$, where $c_{1}$, $h_{1}$, and $w_{1}$ represent dimension, height and width of the extracted feature map. 
    Following semantic extractor, channel encoder is deployed to reduce the number of symbols for low communication overhead, whose output is the broadcast signal $\bm{z} \in \mathbb{R}^{c_{tx}}$, where $c_{tx}$ denotes the number of transmitted symbols.
    Notably, the reduction of symbols is also represented as feature compression.
    The function of feature compression is denoted as $E_{C}::\mathbb{R}^{c_{1} \times h_{1} \times w_{1}} \to \mathbb{R}^{c_{tx}}$.  
    In short, the encoding process in a broadcast semantic communication system can be given by
    \begin{equation}
        \label{SEMD}
		\bm{z} = E_{C}\left(E_{S}\left(\bm{x}; \bm{\phi}_{S}\right); \bm{\phi}_{C}\right),
	\end{equation}
    where $\bm{\phi}_{S}$ and $\bm{\phi}_{C}$ denote optimizable parameters of $E_{S}$ and $E_{C}$ respectively. 
    As the encoding process is done, transmitter antennas broadcasts $\bm{z}$ through heterogeneous physical channel to multiple users. 
\subsubsection{Physical Channel}
    The received broadcast signal for user $i$ is given by
	\begin{equation}
		\bm{\hat{z}}_i = \bm{h}_{i}\bm{z}+\bm{n}_i, \label{eq3}
	\end{equation}
	where the $\bm{h}_i$ denotes the channel gain of user $i$ and $\bm{n}_i$ represents the probability distribution of the $i$-th physical channel. 
	In this paper, we mainly consider three types of physical channel generally existing in real world, which are the AWGN channel, Rayleigh fading channel, and Rician fading channel. 
	As for the AWGN channel, the channel gain $\bm{h}$ becomes $1$ and the distribution $\bm{n}$ follows the Gaussian distribution $(0, \sigma^2\bm{I})$. 
	The channel gain $\bm{h}$ of the Rayleigh fading channel follows the complex Gaussian distribution $\mathcal{CN}\left(0,\bm{I}\right)$ and the noise distribution $\bm{n}$ of the Rayleigh fading channel is similar to that of the AWGN channel. 
	If the physical channel is the Rician fading channel, the channel gain $\bm{h}$ follows the complex Gaussian distribution $\mathcal{CN}\left(\mu_{Rician}, \sigma_{Rician}\bm{I}\right)$ and $\bm{n}$ follows the Gaussian distribution $(0, \sigma^2\bm{I})$.
	The $\mu_{Rician}$ and $\sigma_{Rician}$ are given by 
	\begin{subequations}
		\begin{align}
			\mu_{Rician} = \sqrt{a/\left(a+1\right)},\\
			\sigma_{Rician} = \sqrt{1/\left(a+1\right)}, 
 		\end{align}
	\end{subequations}
	where the $a$ is the Rician coefficient. A higher $a$ denotes more severe channel condition.

\subsubsection{Receivers}
    As the broadcast signal corrupted by physical noise is received, each receiver processes it with multiple NNs and semantic executor. 
    In each receiver, semantic executor aims at accurately executing specific task based on decoded signal. 
    In this paper, we consider multiple image classification tasks mainly. 
    Thus, we define $i$-th semantic decoding function as $D_{S_i}::\mathbb{R}^{c_{tx}} \to \mathbb{R}^{n_{label}}$ respectively, where $n_{label}$ denotes the number of classes of $i$-th classification task. 
    Briefly, the output of $i$-th receiver can be represented as 
    \begin{equation}
        \bm{\hat{y}}_{i} = D_{S_i}\Big(D_{C_i}(\bm{\hat{z}}_i;\bm{\theta}_{C_i});\bm{\theta}_{S_i} \Big), \label{eq4}
    \end{equation}
    where $\bm{\theta}_{C_i}$ and $\bm{\theta}_{S_i}$ denote the optimizable parameters of $D_{C_i}$ and $D_{S_i}$ respectively.
    
\subsection{Problem Description}
    Previous works, such as \cite{sagduyu2023multi}, achieved feature compression by following the configurations widely applied in one-to-one systems \cite{shao2021learning, sheng2022multi, zhang2022unified}. 
    Generally, serial architectures such as stacked fully connected layers (FCs) shown in Fig.~\ref{fig:SEMD} are deployed for feature compression. 
    However, shifting from unicast to broadcast, transmitter meets challenges in encoding. 
    
    As the number of users increases, transmitter needs to keep communication overhead low while containing semantic information for all users.
    In practical, in a broadcast system, the channel encoder must simultaneously accomplish the compression and encoding of task-relevant information for each task, while also being able to handle the fusion of broadcast signals towards heterogeneous channels. 

\subsection{Solutions}
    To tackle the problems, we propose task-channel-aware sub-encoders to parallelly encode semantic information towards downstream task and physical channel for each user.
    Moreover, we develop channel-aware feature fusion sub-encoder to effectively merge user-specific semantic features into broadcast signal according to general channel condition. 
    As we mentioned in Section.~\ref{relatedworks}, it is not easy to achieve fusion for broadcast signal well. 
    Accordingly, previous end-to-end training approaches are no longer efficient enough to realize the full potential of the system.
    To sufficiently optimize the proposed task-oriented broadcast semantic communication system, we leverage the advantage of IB in feature compression. 
    In this paper, we define the position of bottleneck in the proposed broadcast system and design an objective function based on IB theory.

\section{Methodology, Neural Network Design, And Training Strategy} \label{proposedmethod}
    In this section, we will introduce our approach for designing DeepBroadcast and describe our training strategy both in details. 
    \begin{figure}[!t]
        \centering
        \includegraphics[scale=0.45]{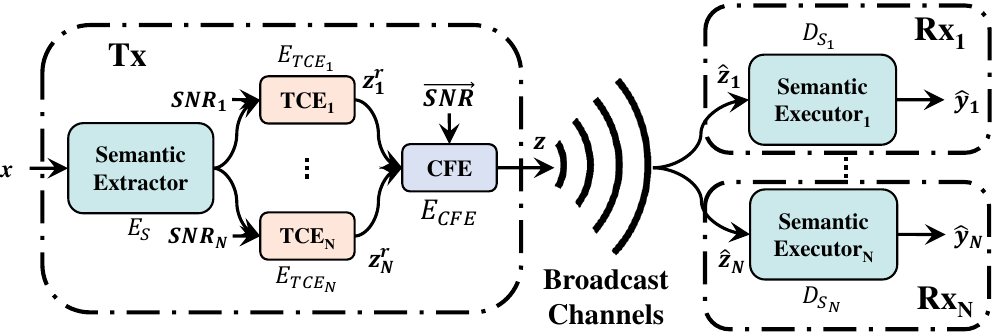}
        \caption{The proposed semantic communication system model.}
        \label{fig:systemModel}
    \end{figure}
    \begin{figure*}[!t]
        \centering
        \includegraphics[scale=0.45]{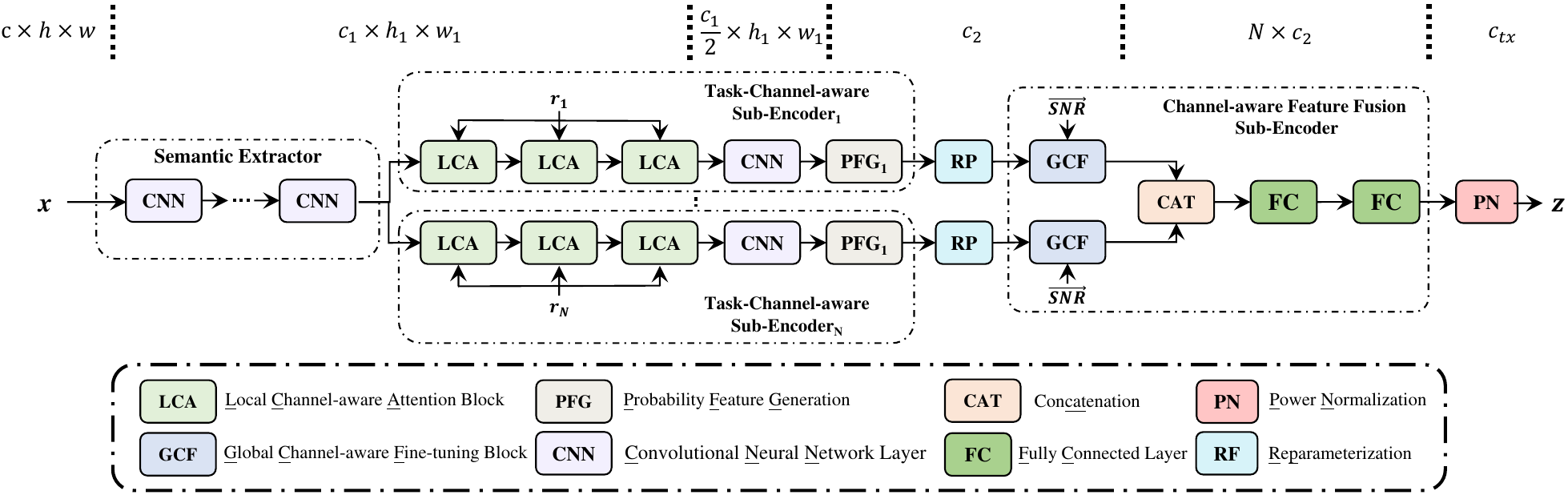}
        \caption{The pipeline of the transmitter of DeepBroadcast.}
        \label{fig:pipeline}
    \end{figure*}
\subsection{Methodology}
    DeepBroadcast is constructed based on deep neural network structutures. 
    As shown in Fig.~\ref{fig:systemModel}, transmitter of DeepBroadcast consists of a semantic extractor, multiple parallel task-channel-aware sub-encoders (TCEs), and a channel-aware feature fusion sub-encoder (CFE). 
    Meanwhile, each receiver is primarily includes a semantic executor. 
    In the following, we will rewrite encoding process function showing in (\ref{SEMD}) to adapt to proposed architecture of DeepBroadcast. 
    
    Recalling the semantic extracting function $E_S$, the TCE function for $i$-th user is given by $E_{TCE_i}::\mathbb{R}^{c_{1} \times h_{1} \times w_{1}} \to \mathbb{R}^{c_{2}}$ where $c_{2}$ represents the number of $i$-th signal dimension after compression. 
    Particularly, we have the equation $c_{2}=\frac{1}{4}\times c_{1} \times h_{1} \times w_{1}$, where the semantic feature extracted by semantic extractor is highly compressed by $i$-th TCE.
    To adapt semantic feature to i-th physical channel, $E_{TCE_i}$ also incorporates $i$-th CSI $r_i$.
    Moreover, in order to leverage variational approach to train the encoders, we employ the stochastic encoder. 
    In particular, $E_{TCE_i}$ generates the distribution parameters of the semantic features, ie., $\bm{\mu}_i \in \mathbb{R}^{c_{2}}$ and $\bm{\sigma}_{i} \in \mathbb{R}^{c_{2}}$.
	Therefore, the outputs of $i$-th TCE can be represented by 
	\begin{equation}
		\bm{\mu}_i, \bm{\sigma}_i = E_{TCE_i} \Big(E_{S} \left(\bm{x};\bm{\phi}_{S}\right), r_i;\bm{\phi}_{TCE_i} \Big), \label{eq1}
	\end{equation}
    where $\bm{\phi_{S}}$ and $\bm{\phi_{TCE_i}}$ denote optimizable parameters of semantic extractor and $i$-th TCE, respectively. 
    In particular, $\bm{\mu}_i$ and $\bm{\sigma}_i$ are treated as mean value and standard deviation of Gaussian distribution for reparameterization. 
	Following \cite{shao2022task}, we adopt the reparameterization trick to sample $\bm{z}^{r}_{i}$ from the learned distribution to optimize the negative log-likelihood term in the proposed loss function (\ref{broadcastIB}). 
    Accordingly, we have 
    \begin{equation}
        \label{reparameterization}
		\bm{z}_{i}^{r} = \bm{\mu}_i + \bm{\sigma}_i \odot \bm{\lambda}_{i},
	\end{equation}
    where $\bm{\lambda}_{i}$ is sampled from $\mathcal{N}(0,1)$. 
	Simply concatenating all refined semantic features into broadcast signal not only significantly increase the communication overhead but also brings redundancy of shared semantic information. 
    Alternatively, we propose CFE to not only decrease number of transmitted symbols but also flexibily adapt broadacast signal to general channel condition in this paper.
    Following TCEs, the CFE aims at merging shared semantic information, task-specific information, and information of global channel condition into the broadcast signal. 

    Overall, taking refined semantic features $\bm{z}^r=[\bm{z}^r_1, \cdots, \bm{z}^r_N]$ and CSI of all channels as inputs and outputing broadcast signal $\bm{z} \in \mathbb{R}^{c_{tx}}$, the function of CFE can be written as 
	\begin{equation}
		\bm{z} = E_{CFE}(\bm{z}^r,\overrightarrow{\bm{SNR}}; \bm{\phi}_{CFE}),
	\end{equation}
	where $\bm{\phi}_{CFE}$ and $\overrightarrow{\bm{SNR}}$ denote optimizable parameter of CFE and global broadcast CSI, respectively.
    Particularly, the CFE is composed of $N$ global channel-aware fine-tuning blocks (GCFs) for $N$ users, a concatenation layer, and a fusion layer. 
	For $i$-th user, the GCF aims at jointly fine-tuning $\bm{z}^{r}_i$ based the broadcast CSI $\overrightarrow{\bm{SNR}}$, which includes CSI of all physical channels, to generate $\bm{z}^{r'}_i$.
	
    As all fine-tuned semantic features $[\bm{z}^{f}_1, \cdots, \bm{z}^{f}_N]$ are generated, the concatenation layer unites them into one signal $\bm{z}_{CAT} \in \mathbb{R}^{N \times c_{2}}$.
	Subsequently, fusion layer reduces the number of symbols in $\bm{z}_{CAT}$ with two FCs and the fused signal is power normalized for controling communication overhead in power level.
    Generally, assuming signal $\bm{s}$ is the input, the process of a FC is given by 
    \begin{equation}
        \label{FC}
        \mathrm{FC}(\bm{s}) = \bm{W} \otimes \bm{s} + \bm{b},
    \end{equation}
    where $\bm{W}$ and $\bm{b}$ represent optimizable weight matrix and bias matrix respectively. 
    Here, ``$\otimes$'' denotes matrix multiplication. 
    Moreover, following \cite{shen2020powernorm}, power normalization is given by 
	\begin{equation}
        \label{powerNorm}
		\mathrm{PowerNorm}(\bm{s}) = \frac{\bm{s}}{\sqrt{\frac{1}{L} \sum_{j=1}^{L}s_{j}^{2}}},
	\end{equation}
	where $\bm{s}$ denotes the input signal.
	Briefly, the broadcast signal $\bm{z}$ is given by 
    \begin{equation}
        \label{fusion}
		\bm{z} = \mathrm{PowerNorm}\left(\mathrm{FC}(\mathrm{FC}(\bm{z}_{CAT}))\right).
	\end{equation}

    Besides the rewrited encoding process shown in above, other parts of DeepBroadcast have been presented in Section~\ref{systemmodel}.

\subsection{Neural Network Design}
\subsubsection{Semantic Extractor}
    Instead of optimizing the backbone for better semantic feature extraction, we concern more about how to achieve task-channel-aware feature compression and fusion in channel encoder in this paper. 
    Accordingly, following some previous works such as \cite{shao2022task,sagduyu2023multi}, we employ multiple CNN blocks in semantic extractor to deeply extract semantic information from source information. 

\subsubsection{Task-Channel-aware Sub-Encoder}
    Without loss of generality, we develop TCEs to adapt semantic information to broadcast channels.
    In particular, each TCE includes three local channel-aware attention blocks (LCAs), a CNN layer, and a Probability Feature Generation (PFG). 
    With extracted semantic feature and information of $i$-th channel conditon, $i$-th TCE outputs the probability features $\bm{\mu}_{i}$ and $\bm{\sigma}_{i}$.
    Notably, we develop LCAs to preserve semantic information from noise distortion by adjusting amplitudes of extracted semantic features according to particular CSI. 
    Moreover, we adopt a CNN layer and PFG module to reduce number of symboles of semantic features in two stages. 

    \begin{figure*}[!t]
        \centering
        \includegraphics[scale=0.46]{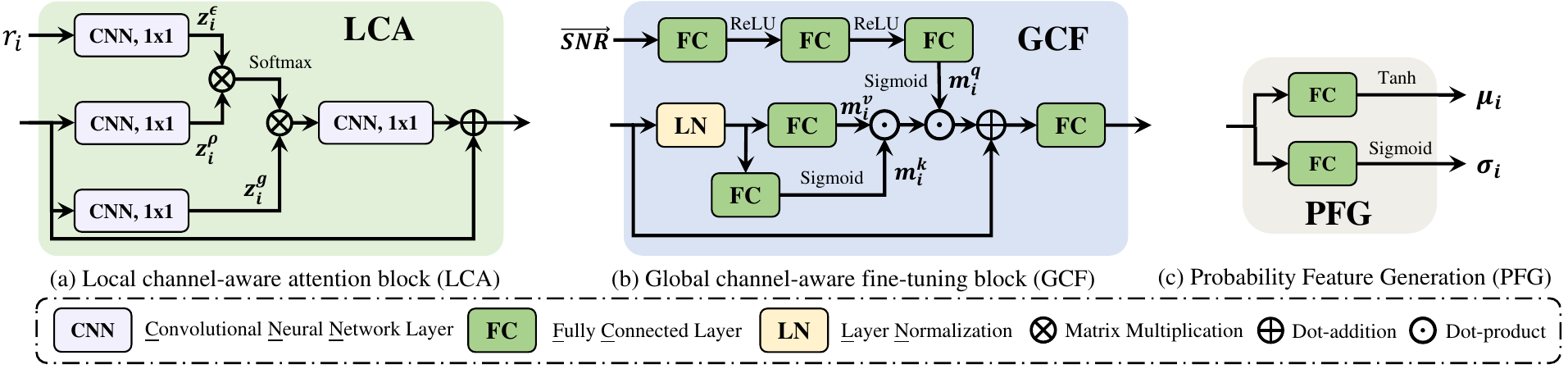}
        \caption{The illustrations of the proposed local channel-aware attention module and global channel-aware fine-tuning module}
        \label{fig:components}
    \end{figure*}

    As shown in Fig.\ref{fig:components}, $i$-th LCA takes certain CSI and semantic feature compressed by FCs as inputs and outputs semantic feature which is aware of $i$-th channel condition. 
    Inspired by \cite{wang2018non}, LCA consists of four CNN layers providing $1 \times 1$ convolutions.  
    Taking the first one of the LCA for $i$-th user as an example for simplicity, the LCA requires extracted semantic features, denoted as $\bm{z}^e \in \mathbb{R}^{c_{1} \times h_{1} \times w_{1}}$, and CSI \big(e.g., signal-noise-ratio (SNR)\big) of the $i$-th physical channel as inputs and outputs the altered semantic features.
    Respectively, $\bm{z}^s$ is processed by convolution kernels to generate $\bm{z}^{\rho}_{i} \in \mathbb{R}^{c_{1} \times h_{1} \times w_{1}}$ and $\bm{z}^{g}_{i} \in \mathbb{R}^{c_{1} \times h_{1} \times w_{1}}$ and $r_i$ is utilized to generate $\bm{z}^{\epsilon}_{i} \in \mathbb{R}^{c_{1} \times h_{1} \times w_{1}}$.
    Then $\bm{z}^{\epsilon}_{i}$ and $\bm{z}^{\rho}_{i}$ are multiplied by each other and the multiplication result is further processed by softmax\footnote{https://en.wikipedia.org/wiki/Softmax\_function} computation. 
    The computation result will be multiplied by $\bm{z}_{i}^{g}$ and then convolution operation is performed on the new matrix. 
    In the following, we leverage element-wise addition to sum the convoluted matrix and $\bm{z}^e$ to obtain the channel-aware semantic features. 

    Following the last LCA, a CNN layer is employ for reducing half of the number of symbols in channel-aware semantic features.
    Taking the partially compressed features as input, $i$-th PFG is developed to generate the probability features $\bm{\mu}_{i}$ and $\bm{\sigma}_{i}$. 
    As shown in Fig.\ref{fig:components}(c), two FC are respectively applied to obtain probability features $\bm{\mu}_i$ and $\bm{\sigma}_i$. 
    According to (\ref{reparameterization}), $\bm{\mu}_{i}$ and $\bm{\sigma}_{i}$ are utilized to generate $\bm{z}^{r}_{i}$, which is one of the inputs of CFE. 

\subsubsection{Channel-aware Feature Fusion Sub-Encoder}
    In this paper, we propose the channel-aware feature fusion sub-encoder to generate broadcast signal and adapt it to heterogeneous broadcast channels.
    The sub-encoder contains a GCF, a concatenation layer (CAT), and a fusion layer.
    As shown in Fig.\ref{fig:pipeline} and Fig.\ref{fig:components}, with inputs of $N$ refined semantic features and information of general channel condition, the sub-encoder outputs the broadcast signal $\bm{z}$. 
    
    Taking $i$-th refined semantic feature $\bm{z}^r_i$ and CSI of all physical channels $\overrightarrow{\bm{SNR}}$ as inputs, the GCF generates $\bm{z}^f_i$. 
    Basically, the GCF consists of multiple FCs. 
    In the early process of GCF, $\bm{z}^r_i$ is normalsized by a layer normalization layer, denoted as $\mathrm{LN}$, and processed self-key modulation and self-value modulation to generated modulated features $\bm{m}_{i}^{k} \in \mathbb{R}^{c_3}$ and $\bm{m}_{i}^{v} \in \mathbb{R}^{c_3}$ respectively. 
    Meanwhile, channel-query modulation transforms CSI input $\overrightarrow{\bm{SNR}}$ into a vector $\bm{m}_{i}^{q} \in \mathbb{R}^{c_3}$. 
    These modulation functions consist of FCs and activation functions including Rectified Linear Unit (ReLU)\footnote{https://en.wikipedia.org/wiki/Rectifier\_(neural\_networks)} and Sigmoid\footnote{https://en.wikipedia.org/wiki/Sigmoid\_function}. 
    Briefly, the outputs of self-key and self-value modulation functions are given by 
    \begin{subequations}
		\label{GCF_kv}
		\begin{align}
        \bm{m}^{k}_{i} &= \mathrm{Sigmoid}\Big(\mathrm{FC}(\mathrm{LN}\left(\bm{z}^{r}_{i}\right))\Big), \label{GCF_kvA}\\
        \bm{m}^{v}_{i} &= \mathrm{FC}(\mathrm{LN}\left(\bm{z}^{r}_{i}\right)), \label{GCF_kvB}
        \end{align}
	\end{subequations}
    and the channel-query modulation function is given by 
    \begin{subequations}
		\label{query2}
		\begin{align}
            \bm{m}^{q}_{(i,1)} = \mathrm{ReLU}\Big(\mathrm{FC}(\overrightarrow{\bm{SNR}})\Big), \\
            \bm{m}^{q}_{(i,2)} = \mathrm{ReLU}\Big(\mathrm{FC}(\bm{m}^{q}_{(i,1)})\Big), \\
            \bm{m}^{q}_{i} = \mathrm{Sigmoid}\Big(\mathrm{FC}(\bm{m}^{q}_{(i,2)})\Big).
        \end{align}
	\end{subequations}
    Then, $\bm{m}^{v}_{i}$ is sequentially dot-multiplying with $\bm{m}^{k}_{i}$ and $\bm{m}^{q}_{i}$ and the result is dot-added with $\bm{z}_{i}^{r}$ to reuse semantic information as well as benefit the backpropagation during training. 
    In the end, the dot-added results are processed by a FC to generate the fine-tuned semantic features $\bm{z}_{i}^{f}$. 
    
    Following GCFs, the CAT merges $N$ fine-tuned semantic features $[\bm{z}^f_{1}, \cdots, \bm{z}^f_{N}]$ together into $\bm{z}_{CAT}$. 
    As shown in (\ref{fusion}), following the CAT, a FC is employed in fusion layer to further compress $\bm{z}_{CAT}$ into broadcast signal $\bm{z}$ to reduce transmission overhead.
    Furthermore, we adopt power normalization (\ref{powerNorm}) to adjust the amplitudes of the broadcast signal to reduce the communication overhead in energy level. 
   
    With the help of CFE, the transmitter can adapt broadcast signal to global channel condition with high efficiency. 
    In Section~\ref{experiment}, we will show the superiority of the proposed sub-encoders.

\subsubsection{Semantic Executor}
    As we mentioned in Section~\ref{systemmodel}, we mainly consider image classification tasks in this paper.
    Similarly, we follow the general paradigm \cite{shao2021learning, shao2022task,sagduyu2023multi} to deploy multiple FCs in each semantic executor for corresponding classification task. 

    Overall, DeepBroadcast is constructed based on neural networks mentioned in this section. 
    Compared with previous approaches, we particularly design novel structures to adapt broadcast signal to heterogeneous channels. 
    However, these structures are also more complex to be trained by previous end-to-end approaches. 
    To realize the full potential of our model, in this paper, we develop an objective function for broadcast semantic communication system. 

\subsection{Training Strategy}
    Based on \cite{alemi2016deep}, the objective function of broadcast semantic communication system consists of terms for accuracy and overhead respectively.
    Here we respectively define the accuracy term as $\mathcal{L}_{accuracy}$ aiming at maximizing the inference accuracy and the overhead term as $\mathcal{L}_{overhead}$ aiming at minimizing communication overhead.
    Consequently, the objective function of broadcast semantic communication system can be briefly represented as 
    \begin{equation}
        \label{broadcastIB_sketchy}
        \mathcal{L}_{Broadcast} = \mathcal{L}_{accuracy} + \beta \mathcal{L}_{overhead},
    \end{equation}
    where $\beta$ is a Lagrange multiplier applied for controling the trade-off between two contradictory objective subfunctions. 

    As we shown in the Appendix, the accuracy term can be optimized by minimizing cross entropy\footnote{https://en.wikipedia.org/wiki/Cross-entropy} objective functions.
    Since each receiver is on duty for a specific task actually, the broadcast task-oriented semantic communication system can be regarded as a special form of multi-task system. 
    As a result, we have 
    \begin{equation}
        \label{broadcastIB_1stterm}
        \mathcal{L}_{accuracy} = \sum_{i=1}^{N} w_{T_i}\mathcal{L}_{T_i}, 
    \end{equation}
    where $w_{T_i}$ represents the task weight of $i$-th device and $\mathcal{L}_{T_i}$ is the loss function of $i$-th task.
    Particularly, $\mathcal{L}_{overhead}$ can be optimized with Kullback-Leibler (KL)-divergence for information reduction.
    Briefly, the second part of extended variational IB objective funtion is formulated as 
    \begin{equation}
        \label{broadcastIB_2ndterm}
        \mathcal{L}_{overhead} = \sum_{i=1}^{N}\gamma_{i}D_{KL}(p_{\phi}(\bm{z}^r_{i}|\bm{x})||q(\bm{z}^r_{i})), 
    \end{equation}
    where $\gamma_{i}$ is a tunable hyperparameter for $i$-th device and for all the $\gamma_{i}$ they obey 
    \begin{equation}
        \sum_{i=1}^{N} \gamma_{i} = 1.
    \end{equation}
    Due to the unpredictable physical channel conditions between transmitter and receivers all the time, it is more general for the broadcast system to set the $\gamma_{i}$ as $\frac{1}{N}$, which makes it fair and unbiased for all the channel conditions as well. 
    Overall, the broadcast IB objective function is given by 
    \begin{equation}
        \label{broadcastIB}
        \begin{aligned}
        \mathcal{L}_{Broadcast}(\bm{\phi}, \bm{\theta}) &= \sum_{i=1}^{N} \Big[w_{T_i}\mathcal{L}_{T_i}\left(\bm{\phi_{S}}, \bm{\phi}_{TCE_i}, \bm{\phi}_{CCF}, \theta_{i}\right) \\ 
                                                        &+\frac{\beta}{N}D_{KL}\big(p_{\bm{\phi}_i}(\bm{z}^r_{i}|\bm{x})||q(\bm{z}^r_{i})\big)\Big],
        \end{aligned}
    \end{equation}
    where $\bm{\theta}_{i}$ represents $\left[\bm{\theta_{S_{i}}}, \bm{\theta_{C_{i}}}\right]$.

    Employing an appropriate training strategy is advantageous for developing the system.
    Algorithm \ref{algorithm1} illustrates the training procedures of our DeepBroadcast. 
    Additionally, the parameter set $\bm{\theta}$ can be given by $\bm{\phi}=\left[\bm{\phi}_S, \bm{\phi}_{TCE_1}, \cdots , \bm{\phi}_{TCE_N}, \bm{\phi}_{CCF}\right]$ while $\bm{\theta}$ is given by $\bm{\theta}=\left[\bm{\theta}_{S_1}, \cdots , \bm{\theta}_{S_N}, \bm{\theta}_{C_1}, \cdots , \bm{\theta}_{C_N}\right]$. 
    For the proposed system, based on the loss function defined in (\ref{broadcastIB}), the model parameters $\bm{\theta}$ and $\bm{\phi}$ can be optimized by minimizing the weighted task-specific loss function values and the KL-divergence. 
    \begin{algorithm}[!t]
        \caption{Training Procedures for DeepBroadcast}
        \label{algorithm1}
        \begin{algorithmic}[1]
            \STATE \textbf{Input}: number of iterations $T$, batch size $B$, number of devices $N$, training SNR list
            \STATE \textbf{while} epoch $t=1$ to $T$ \textbf{do}
            \STATE \hspace{0.5cm} Select a mini-batch of data $\left\{(\bm{x}_b,\bm{y}_b)\right\}^{B}_{b=1}$
            \STATE \hspace{0.5cm} \textbf{while} $i=1$ to $N$ \textbf{do}
            \STATE \hspace{1.0cm} \textbf{Randomly} joint $r_i$ from SNR List to $\overrightarrow{\bm{SNR}}$ for $i$-th simulated physical channel
            \STATE \hspace{0.5cm} \textbf{end while}
            \STATE \hspace{0.5cm} Compute the shared feature vector based on $E_{S}$
            \STATE \hspace{0.5cm} \textbf{while} $i=1$ to $N$ \textbf{do}
            \STATE \hspace{1.0cm} Compute the user-specific probability feature vectors $\left\{\bm{\mu}_{i,b}, \bm{\sigma}_{i,b}\right\}^{B}_{b=1}$ base on (\ref{eq1})
            \STATE \hspace{0.5cm} \textbf{end while}
            \STATE \hspace{0.5cm} Compute the broadcast signal $\left\{\bm{z}_b\right\}^{B}_{b=1}$ base on (\ref{fusion}),(\ref{GCF_kv}), and (\ref{query2}) with $\overrightarrow{\bm{SNR}}$ 
            \STATE \hspace{0.5cm} Compute the overhead loss $\mathcal{L}_{overhead}$ based on (\ref{broadcastIB_2ndterm})
            \STATE \hspace{0.5cm} \textbf{while} $i=1$ to $N$ \textbf{do}
            \STATE \hspace{1.0cm} Compute the noise-corrupted feature vector $\left\{\hat{\bm{z}}_{i,b}\right\}^{B}_{b=1}$ base on (\ref{eq3}) with $r_i$
            \STATE \hspace{1.0cm} Compute the inference result $\left\{\hat{\bm{y}}_{i,b}\right\}^{B}_{b=1}$ based on (\ref{eq4}) and the inference loss $\mathcal{L}_{accuracy}$ based on (\ref{broadcastIB_1stterm})
            \STATE \hspace{0.5cm} \textbf{end while}
            \STATE \hspace{0.5cm} Compute the loss $\mathcal{L}_{Broadcast}(\bm{\theta}, \bm{\phi})$ based on (\ref{broadcastIB})
            \STATE \hspace{0.5cm} Update parameters $\bm{\theta}$, $\bm{\phi}$ through backpropagation
            \STATE \textbf{end while}
        \end{algorithmic}
    \end{algorithm}

    To evaluate the system effectiveness and reliability of the proposed multi-user semantic communication system, several dense prediction tasks, including semantic segmentation, depth estimation, and human parts segmentation etc, are performed in two datasets.
    Both of them include amounts of complex real images with large image size, and more details will be provided in Section~\ref{experiment}.

\section{Performance Evaluation} \label{experiment}
    In this section, we evaluate DeepBroadcast to answer the following questions:
    \begin{itemize}
        \item How does DeepBroadcast achieve adaptation to heterogeneous broadcast channel? (Section~VI.A)
        \item Does DeepBroadcast maintain robust performance across heterogeneous channels and tasks in multi-user scenarios? (Section~VI.B$\&$C)
        \item How does each component and training strategy of DeepBroadcast contribute to the performance gain? (Section~VI.D$\&$E)
    \end{itemize}
    Particularly, we study one case for each question.
    Following previous works such as \cite{bourtsoulatze2019deep, sagduyu2023multi, lyu2024semantic}, all simulations are performed in CIFAR10 dataset for simplicity, consisting 50,000 R.G.B images of size $32 \times 32 \times 3$ for training and 10,000 images for test.
\subsection{Analysis on Visualization Results}
\begin{figure}[!t]
    \centering
    \includegraphics[scale=0.34]{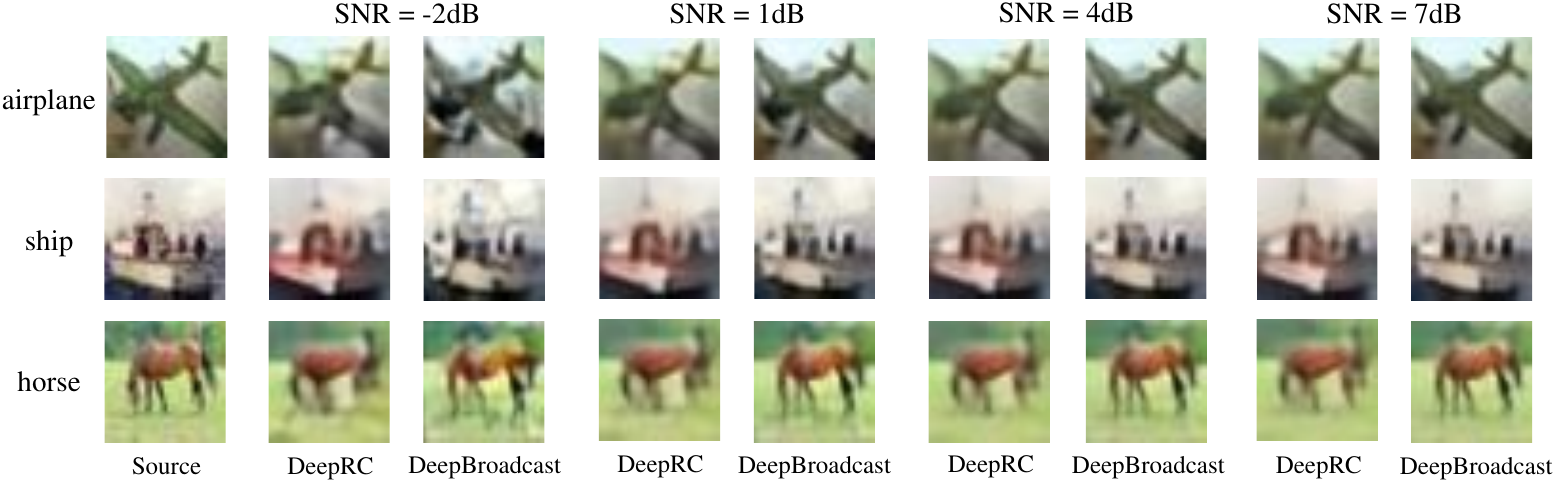}
    \caption{Some visualization results of DeepBroadcast and DeepRC in image recovery in CIFAR10. }
    \label{fig:pre_exp_visualizationIMG}
\end{figure}
\begin{figure*}[!t]
    \centering
    \subfloat[DeepBroadcast/PSNR=26.11dB]{
        \includegraphics[width=0.23\linewidth]{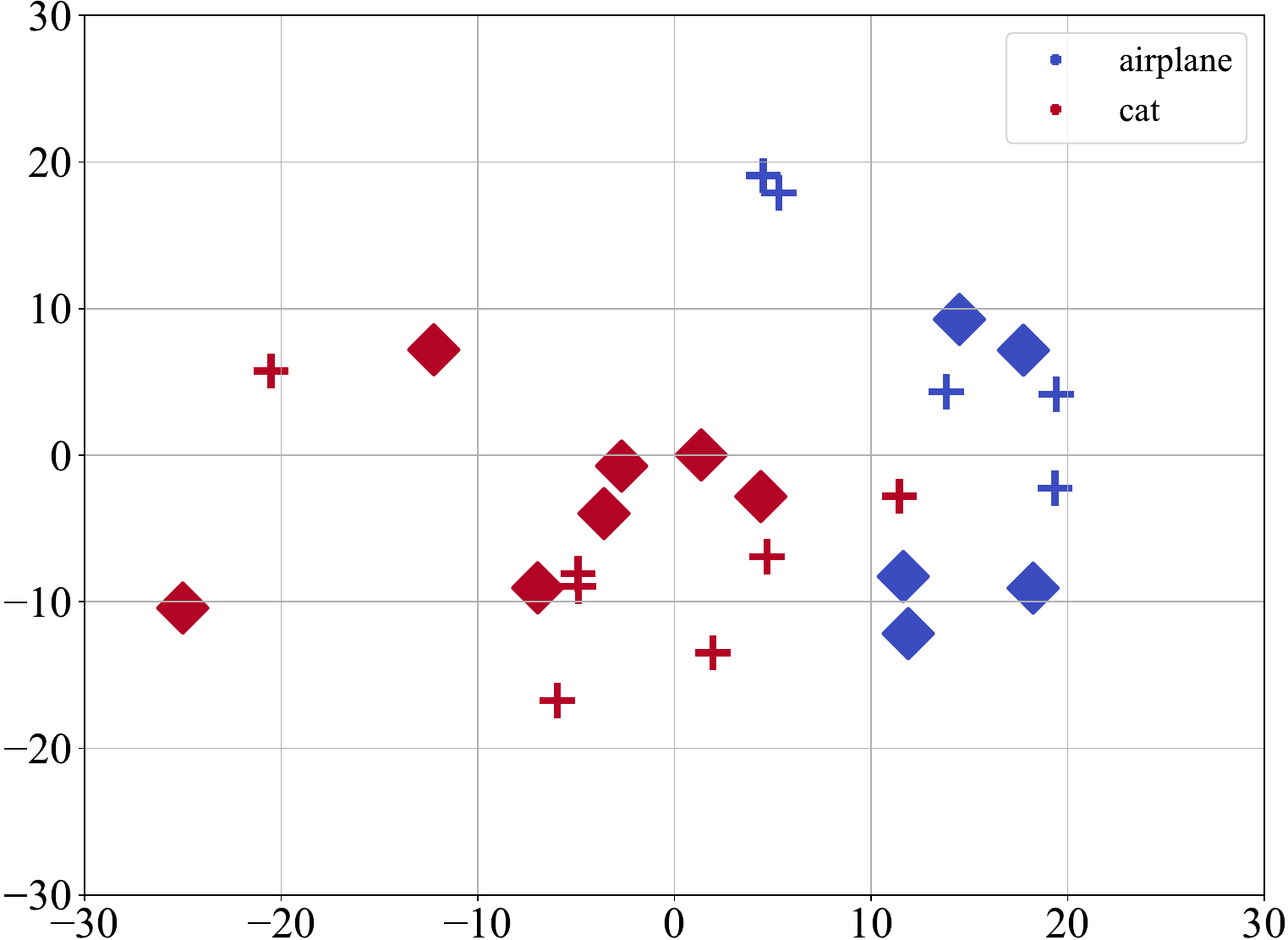}%
        \label{fig:case1_A}}
    \subfloat[DeepRC/PSNR=24.18dB]{
        \includegraphics[width=0.23\linewidth]{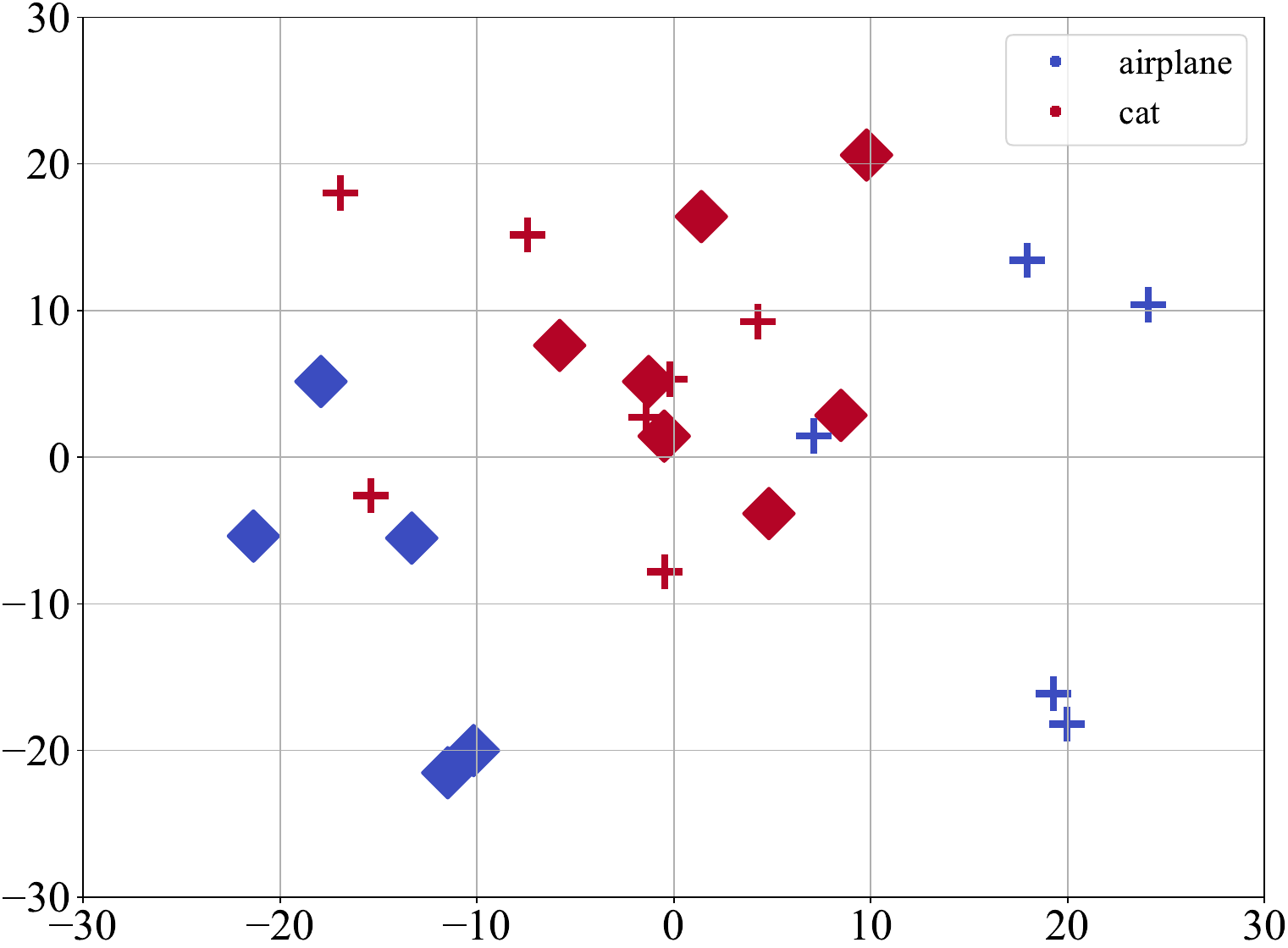}%
        \label{fig:case1_B}}
    \subfloat[DeepBroadcast/Acc=58.27$\%$]{
        \includegraphics[width=0.23\linewidth]{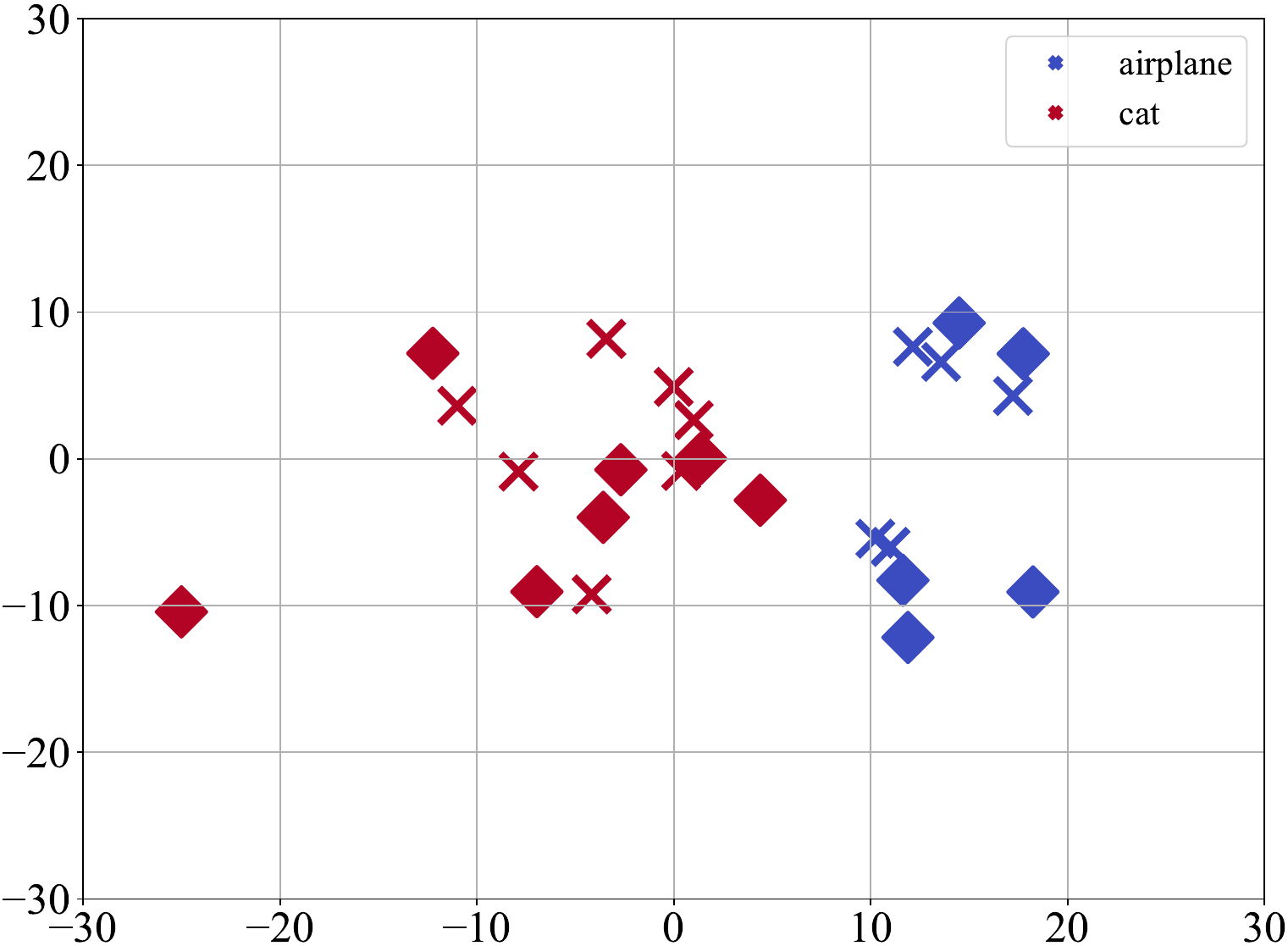}%
        \label{fig:case1_C}}
    \subfloat[DeepRC/Acc=57.32$\%$]{
        \includegraphics[width=0.23\linewidth]{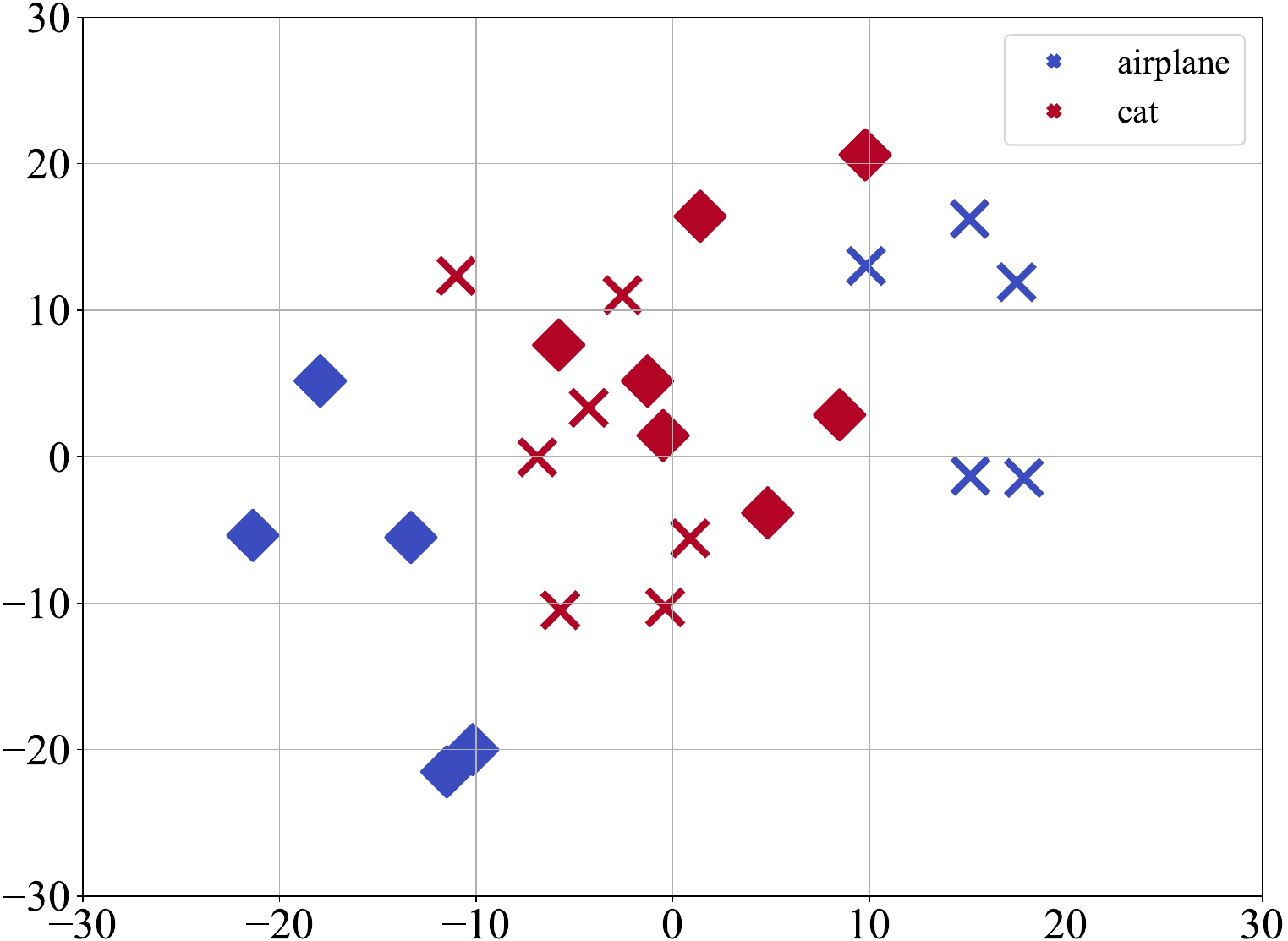}%
        \label{fig:case1_D}}
    \caption{2-dimensional t-SNE embedding of the broadcast signal and received signals in CIFAR10 image recovery and classification task with SNR=7dB. Particularly, blue and red markers denote samples of airplane and cat respectively. The ``$\Diamond$'', ``$+$'', and ``$\times$'' markers represent broadcast signal, received signal of user-1, and received signal of user-2 respectively.}
    \label{fig:case1}
\end{figure*}  

\subsubsection{Setup}
    Inspired by \cite{lyu2024semantic}, we consider a two-user case where the down-steam tasks are image recovery and 10-class image classification, respectively.
    It is worth noting that image recovery is sensible to noise distortion, making it suitable for exploring how the systems achieve noise robustness.
    In case-I, we evaluate DeepBroadcast and the JSCC approach for image recovery and classification (DeepRC), referring the basic network in \cite{lyu2024semantic}.
    For fair comparison, we construct DeepBroadcast and DeepRC with the same backbone and align their depth as much as possible in all experiments.
    Here, we argue that all the source information is important for the image recovery and it is contrary to the purpose we introduced IB.
    Thus, in this case, we follow the common end-to-end training strategy.
    The channel conditions of users are set as AWGN channel and Rayleigh fading channel, respectively. 
    With the similar concern in \cite{lyu2024semantic}, we apply higher weight factor for image recovery since it is with higher difficulty and easier to be corrupted by noise. 
    Thus, we train DeepBroadcast and DeepRC with the following loss function
    \begin{equation}
        \label{LossRC}
		\mathcal{L}_{RC}^{'} = \mathcal{L}_{1} + 10^{-3} \times CE,
	\end{equation}
    where $\mathcal{L}_{1}$ and $CE$ represent L1-loss\footnote{https://en.wikipedia.org/wiki/Least\_absolute\_deviations} and cross entropy respectively.

\subsubsection{Analysis}
\begin{figure}[!t]
    \centering
    \includegraphics[scale=0.2]{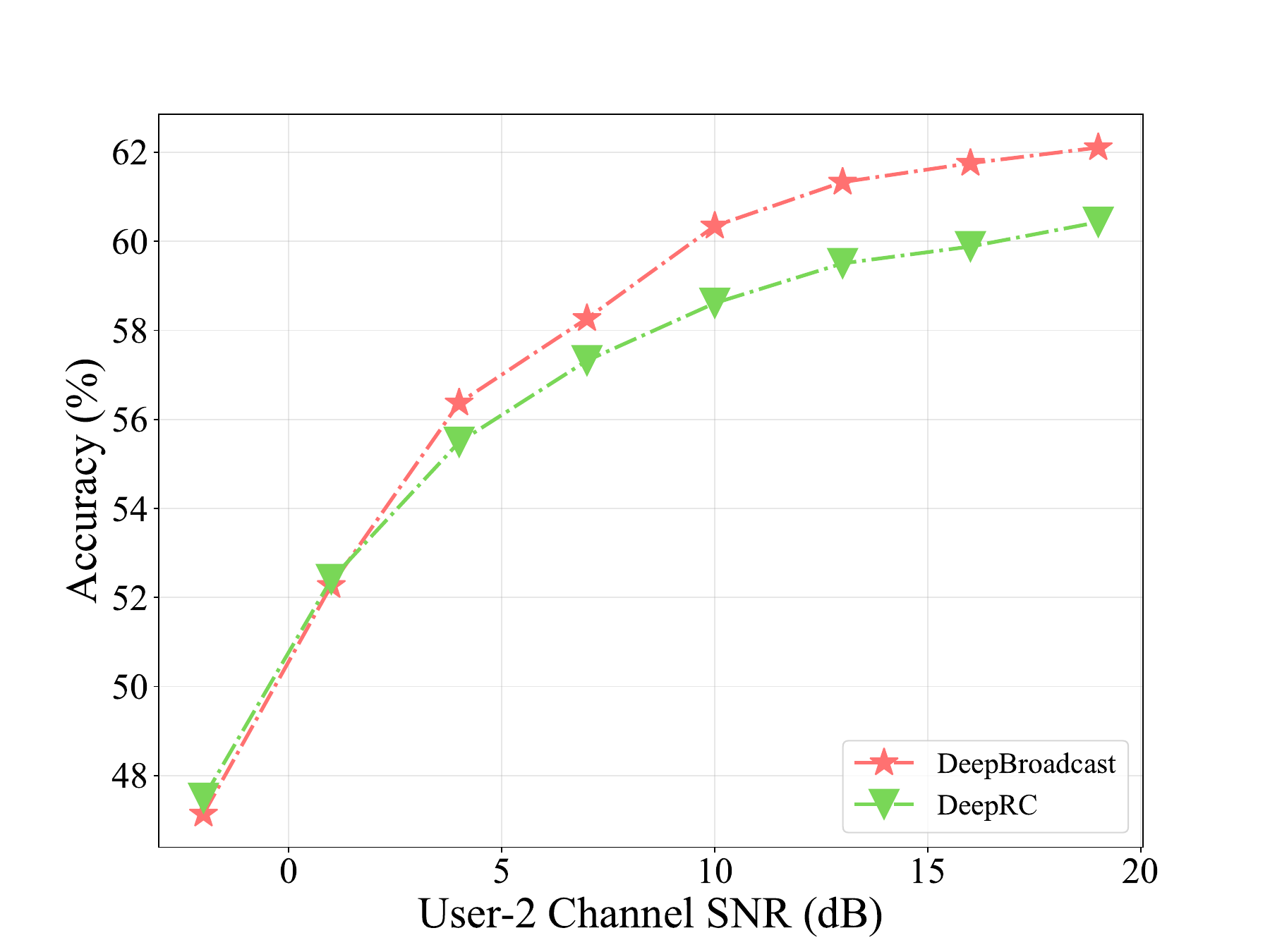}
    \caption{10-class classification accuracy of DeepBroadcast and DeepRC. }
    \label{fig:pre_exp_acc}
\end{figure}

    Fig.~\ref{fig:pre_exp_visualizationIMG} and Fig.~\ref{fig:pre_exp_acc} illustrate simulation results over AWGN channel and Rayleigh fading channel.
    It can be observed in Fig.~\ref{fig:pre_exp_visualizationIMG} that DeepBroadcast achieves comparable visual performance in image recovery over AWGN channel with noise intensity from -2dB to 7dB.
    In addition, as shown in Fig.~\ref{fig:pre_exp_acc}, DeepBroadcast also displays higher classification accuracy than DeepRC.

    In Fig.~\ref{fig:case1}, we provide visualization results of broadcast signal and received signal samples of two users through channels with intensity of 7dB.
    For better analysis, we select samples from two classes and some features from these samples are present.
    Particularly, blue and red markers denote samples of airplane and cat respectively.
    The ``$\Diamond$'', ``$+$'', and ``$\times$'' markers represent broadcast signal, received signal of user-1, and received signal of user-2 respectively.
    In Fig.~\ref{fig:case1}(a) and Fig.~\ref{fig:case1}(b), it can be observed that the clusters of broadcasted signal and received signal are closer in DeepBroadcast, especially those of blue-marked samples from airplane.
    Additionally, the average PSNR of recovered images in DeepBroadcast is 1.93dB higher than that in DeepRC.
    Compared with previous JSCC-based approach DeepRC, both observations demonstrate DeepBroadcast is more robust in the noisy channel. 
    Moreover, although Fig.~\ref{fig:case1}(c) and Fig.~\ref{fig:case1}(d) illustrate the similar results in feature distance, it is worth noting that the achieved classification accuracy gap between DeepBroadcast and DeepRC is 0.95$\%$, which is not as evident as that in recovery.
    Here, we argue that \emph{as long as each region of object class can be divided from the others, the distance between transmitted signal and received signal is not so important}.
    \emph{Moreover, it is the higher degree of freedom in image classification that earns more encoding room to achieve more distinct feature in image recovery. }
    In the following, the advantages of the proposed DeepBroadcast will be emphasized by more simulation results.

\subsection{Performance Comparison in 2-user scenario}

\subsubsection{Setup}

    In case-II, a scenario where two different image classification tasks are charged by two independent down-stream devices is considered. 
    Following \cite{sagduyu2023multi}, Task-1 is to classify images to animals and Task-2 is to classify images to small ground entities.
    In this case, we evaluate DeepBroadcast, task-oriented broadcast system MTOC referring to \cite{sagduyu2023multi}, JSCC-based unicast systems Unicast, and the SSCC scheme BPG and capacity-achieving channel coding approach.
    Particularly, we introduce a wired JSCC model as baseline. 
    The channel conditions of users are set as Rayleigh fading channel and AWGN channel, respectively. 
    Following \cite{zhang2022deep}, it is worth noting that the compression ratio (CR) of image source coding is that of JSCC-based broadcast system times $\frac{1}{2}\log(1+SNR)$.
    Here, CR of JSCC-based broadcast system is given by the division of the number of transmitted symbols and the number of source image pixels in the system.
    The number of transmitted symbols encoded by JSCC-based broadcast system is 16 in all following simulations.
    In particular, we align the communication overhead of broadcast system and unicast systems.
    Here, we allow that of unicast systems be a little higher than that of broadcast system since the number of transmitted symbols must be integer.
    Inspired by multi-task learning approaches, we train the multi-task networks with the following loss function
    \begin{equation}
        \label{Loss_case2}
		\mathcal{L}_{total} = 0.5 \times CE_{1}+0.5 \times CE_{2} + 10^{-4} \times D_{KL}. 
	\end{equation}

    \begin{figure}[!t]
        \centering
        \subfloat[]{
            \includegraphics[width=0.6\linewidth]{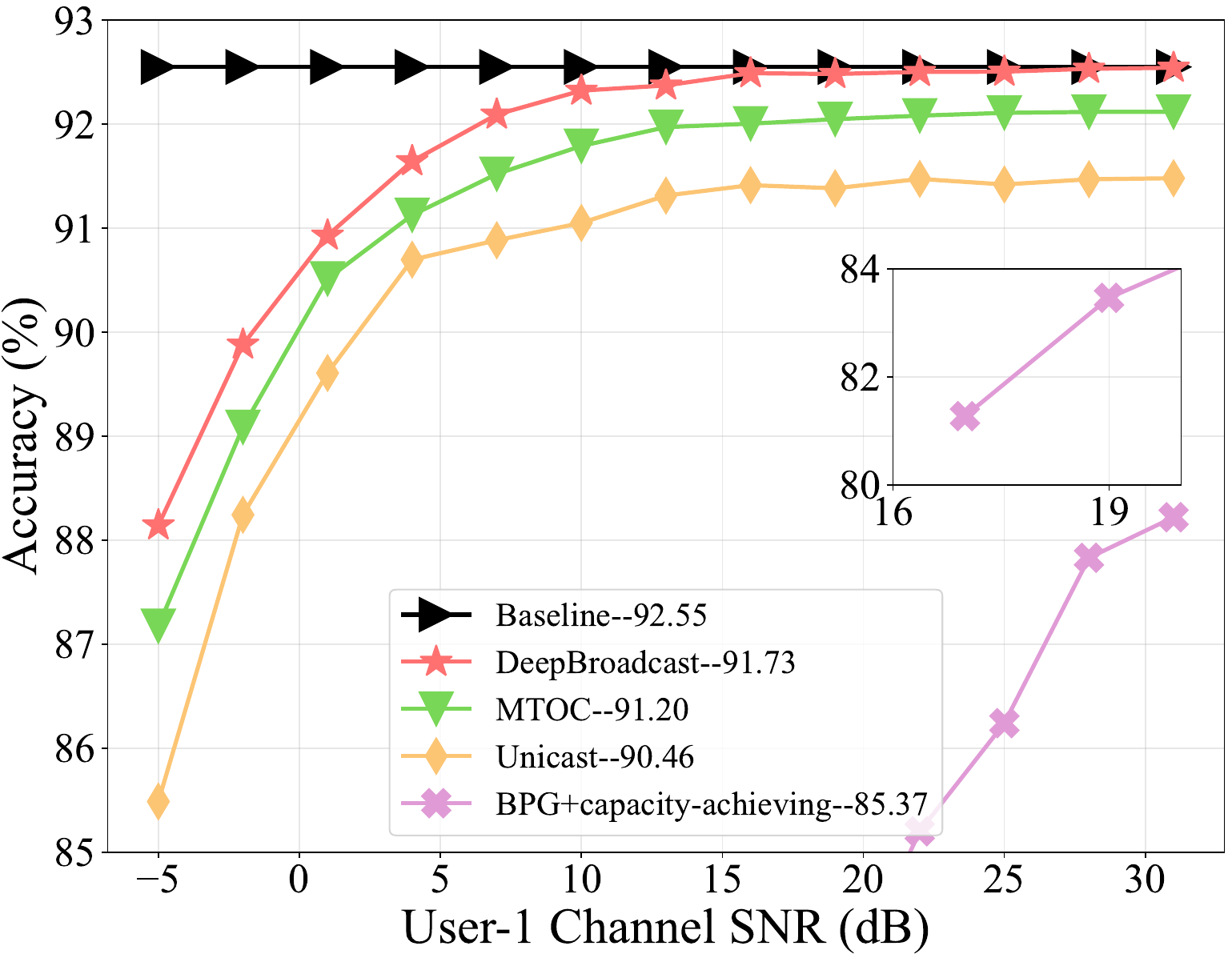}%
            \label{fig:case2_A}} \\ 
        \subfloat[]{
            \includegraphics[width=0.6\linewidth]{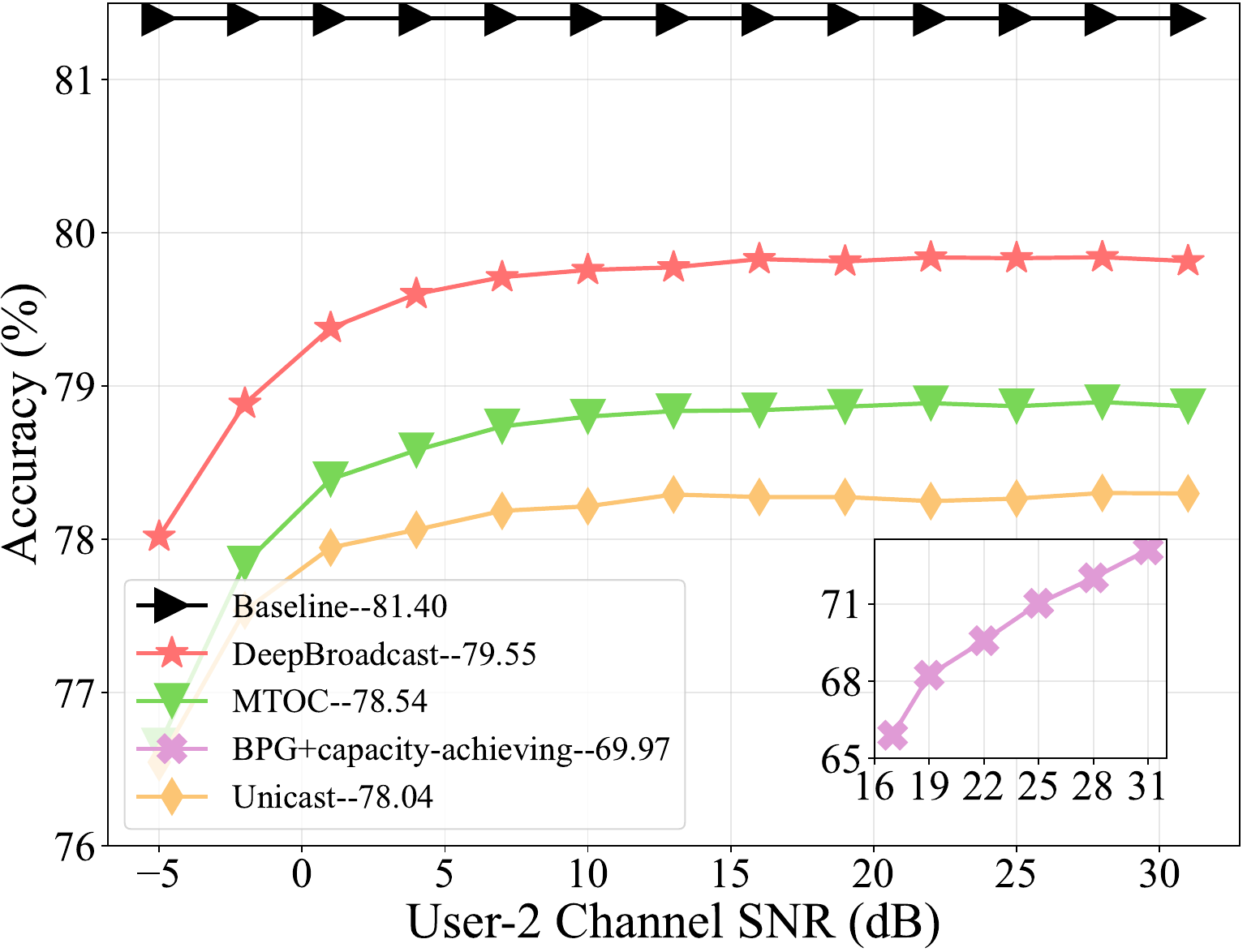}%
            \label{fig:case2_B}}
        \caption{Inference accuracy of 2-user scenario in case-II versus SNR from -5dB to 31dB on CIFAR10 dataset. }
        \label{fig:case2}
    \end{figure}  

\subsubsection{Validation Results}

    In this experiment, we evaluate the average performance achieved by DeepBroadcast and benchmarks across various channel conditions.
    In Fig.~\ref{fig:case2}, we plot the achieved classification accuracy on two tasks and the average accuracy over all considered channel conditions is shown in the legends respectively.
    Compared with Baseline and MTOC, DeepBroadcast displays less performance degradation versus different channel conditions in both tasks.
    When the channel conditions deteriorate, the advantages of the proposed method become more pronounced.
    The proposed DeepBroadcast also outperforms the unicast systems in both tasks.
    As shown by Fig.~\ref{fig:case2}, DeepBroadcast and MTOC both outperform BPG and capacity-achieving channel coding approach.
    In particular, it is worth noting that the communication overhead of JSCC-based systems is so low that the compression ratio of BPG needs to be even higher.
    However, with such communication overhead limitation, BPG fails to compress the CIFAR10 images extremely in a wide range of noise intensity.
    This observation reveals one of the drawbacks of SSCC schemes under the settings including low communication overhead and non-ideal channel condition.
    In the following experiments, we will only evaluate the JSCC-based systems.
    \begin{figure*}[!t]
        \centering
        \subfloat[]{
            \includegraphics[width=0.29\linewidth]{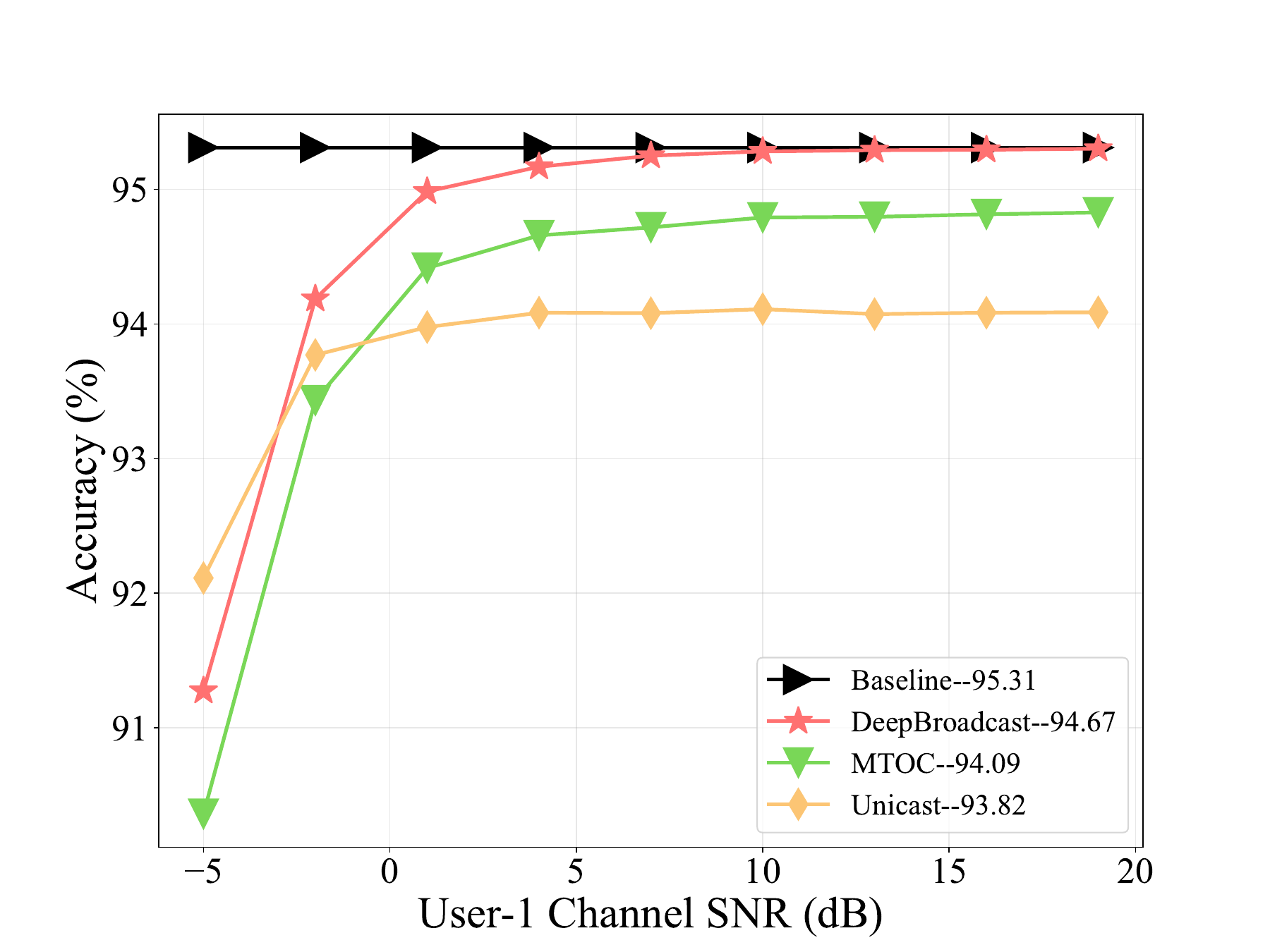}%
            \label{fig:case3_A}}
        \subfloat[]{
            \includegraphics[width=0.29\linewidth]{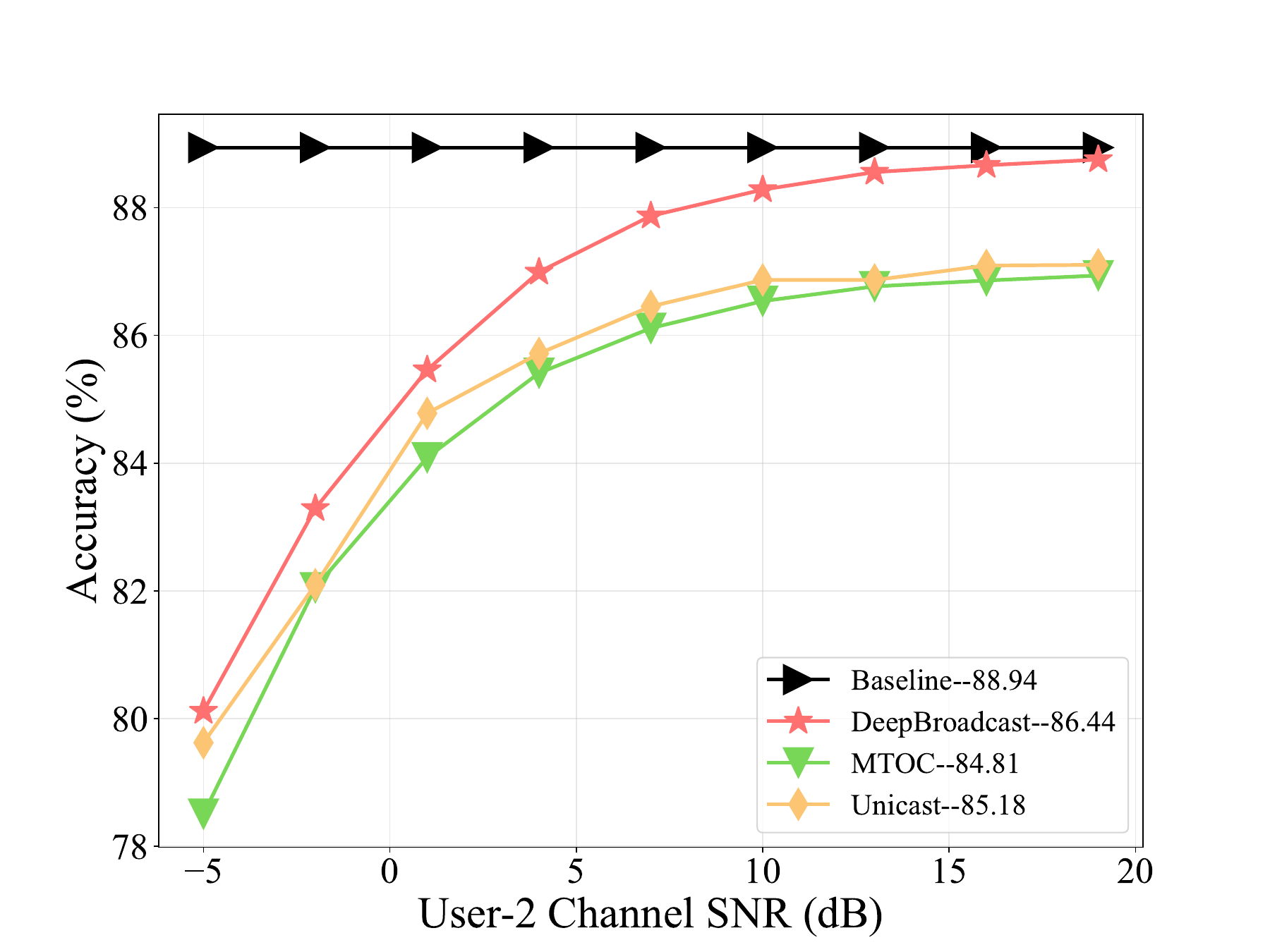}%
            \label{fig:case3_B}}
        \subfloat[]{
            \includegraphics[width=0.29\linewidth]{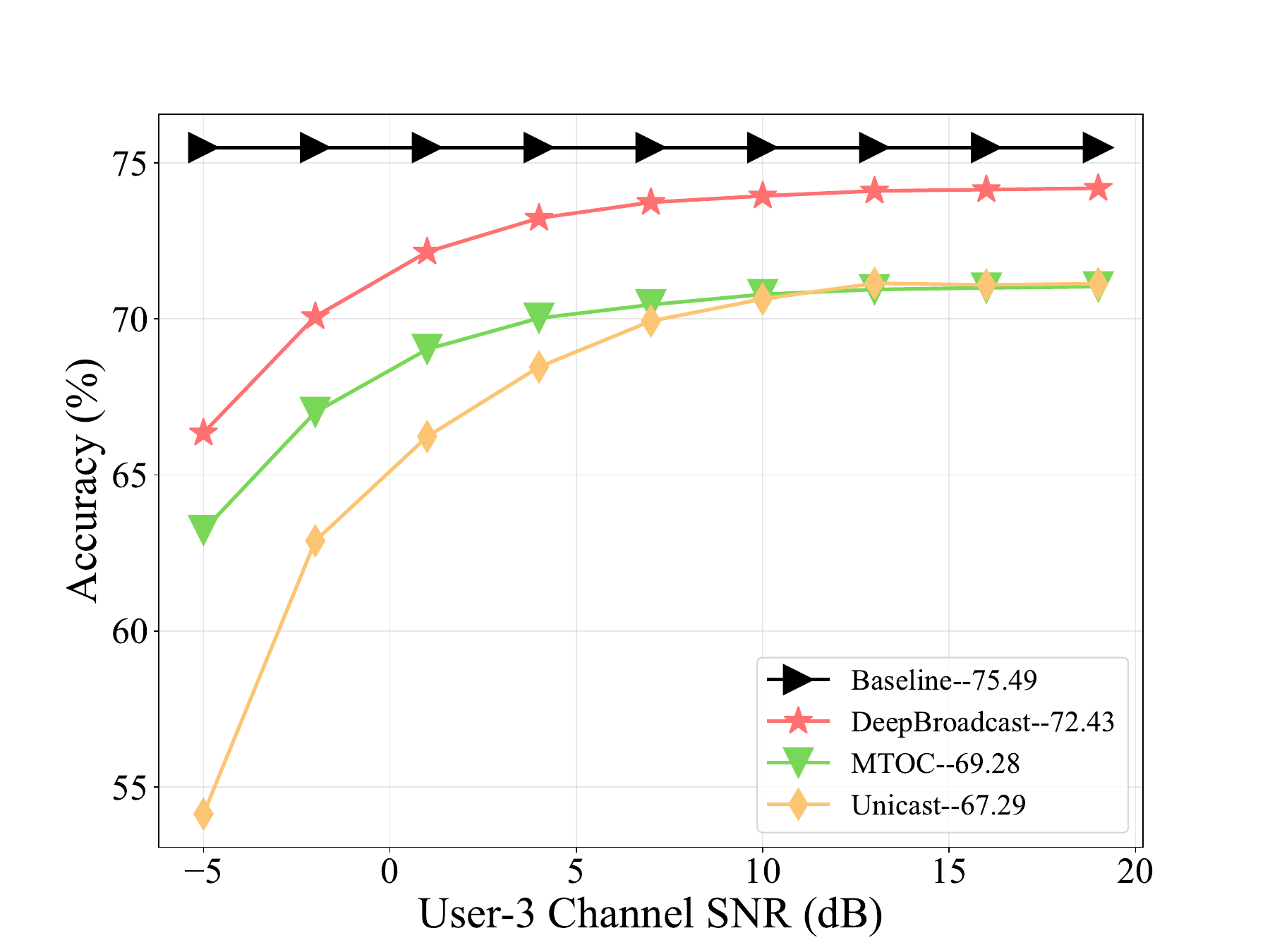}%
            \label{fig:case3_C}}
        \caption{Inference accuracy of 3-user scenario versus SNR from -5dB to 19dB on CIFAR10 dataset. }
        \label{fig:case3}
    \end{figure*}  
    
\subsection{Validation in 3-user scenario}
\subsubsection{Setup} 
  
    In case-III, besides two users and their tasks considered in case-II, we extra introduce a third user who is in charge of 10-class image classification.
    Furthermore, the channels between transmitter and user-1, user-2, user-3 are set as AWGN channel, Rayleigh fading channel, and Rician fading channel respectively.
    Here, the Rician coefficient is set as 2.
    Notably, since we consider Task-3 deserves higher weight for its execution difficulty and severe channel condition, the loss function is updated to
    \begin{equation}
        \label{Loss_case3}
		\mathcal{L}_{total} = 0.15 \times (CE_{1}+CE_{2}) + 0.7 \times CE_{3} + 10^{-4} \times D_{KL}. 
	\end{equation}
    The other implementation settings are similar to previous experiment.
    
\subsubsection{Validation Results}
    As shown in Fig.~\ref{fig:case3}, we present the results of communicating over three heterogeneous channels.
    Briefly, DeepBroadcast achieves better multi-task performance than MTOC in all three tasks.
    Although difference of classification performance of DeepBroadcast and MTOC in Task-1 is not evident, we argue that it is not easy to improve the performance since it is close to perfect inference.
    It is worth noting that DeepBroadcast shows better robustness than MTOC in Task-3 over noisy Rician fading channel.
    Moreover, it can be observed that DeepBroadcast and MTOC significantly outperform Unicast in Task-3 over Rician Fading channel.
    When the channel conditions deteriorate, the advantages of DeepBroadcast become more pronounced. 
    For instance, when the channel SNR of user-3 is $-5$dB, the proposed DeepBroadcast provide additional $3.12\%$ and $12.21\%$ accuracy with MTOC and Unicast, respectively. 

\subsection{Ablation Study on Proposed Components}
\begin{figure}[!t]
    \centering
    \subfloat[]{
        \includegraphics[width=0.6\linewidth]{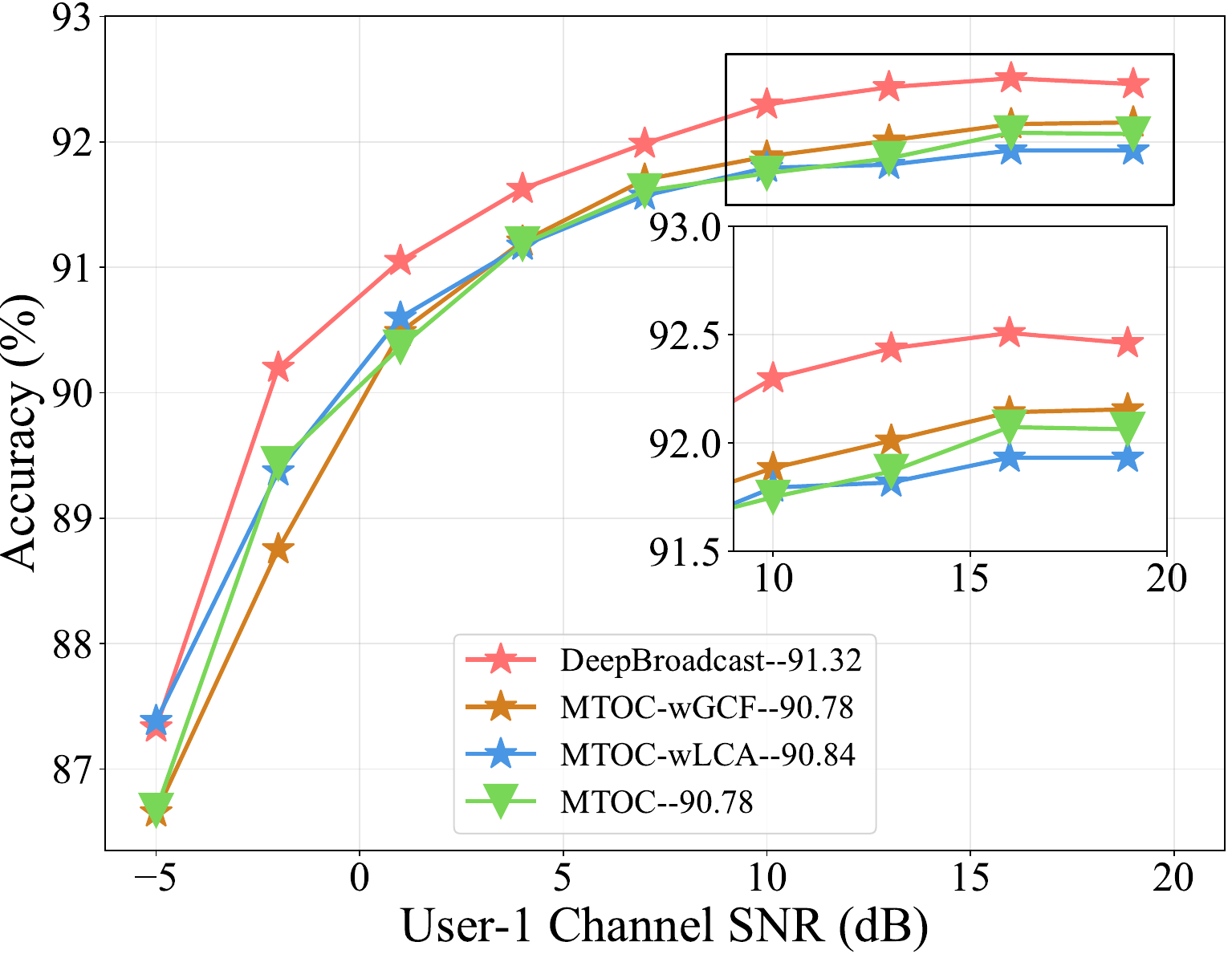}%
        \label{fig:case4_A}} \\
    \subfloat[]{
        \includegraphics[width=0.6\linewidth]{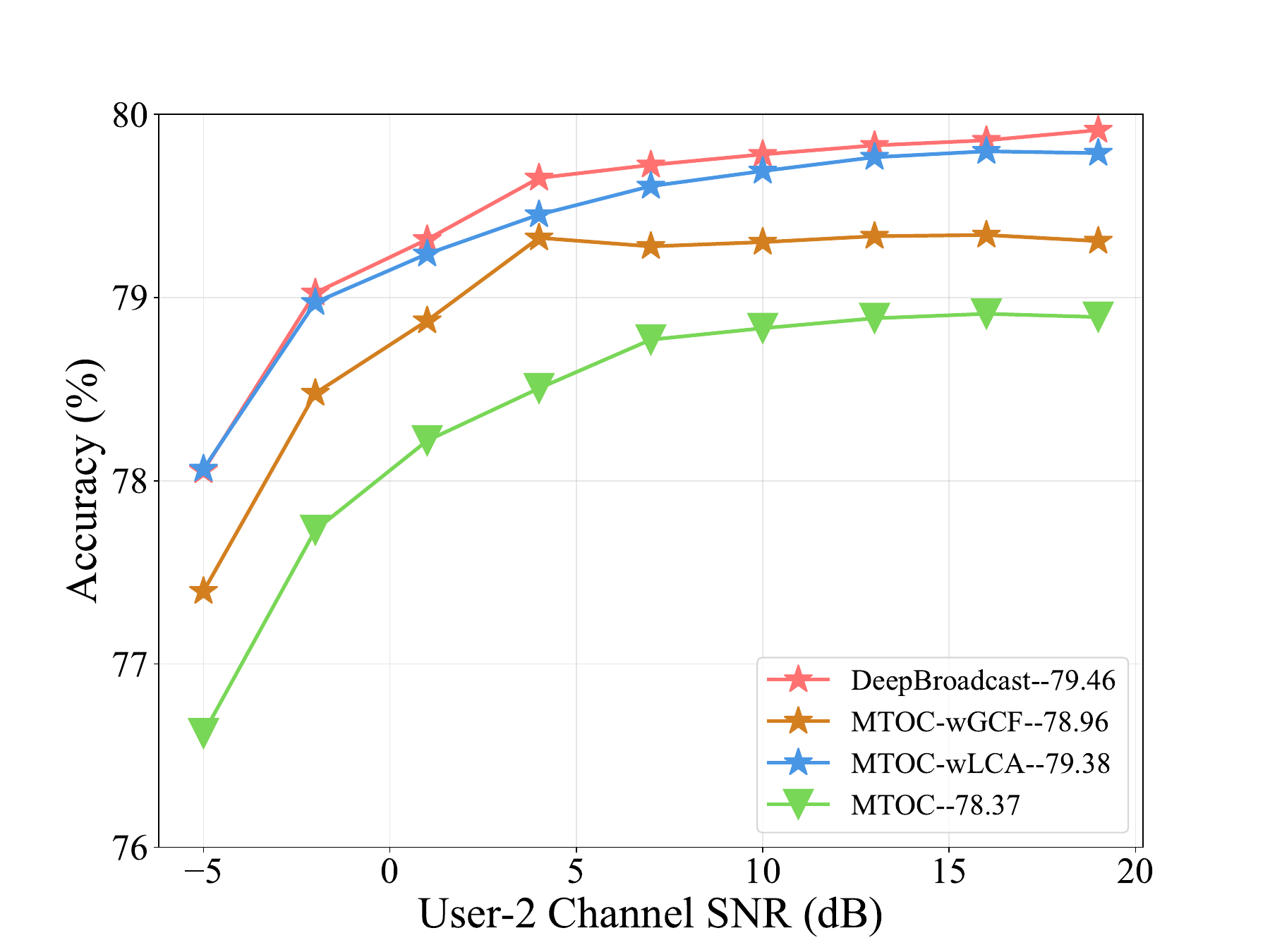}%
        \label{fig:case4_B}}
    \caption{Inference accuracy of 2-user scenario in case-IV versus SNR from -5dB to 19dB on CIFAR10 dataset. }
    \label{fig:case4}
\end{figure}  
\subsubsection{Setup}
    In case-IV, based on the implementation settings in previous experiment of 2-user scenario, we respectively employ LCAs and GCF, as well as sub-encoder architecture, in MTOC to construct MTOC-wLCA and MTOC-wGCF to study the contributions of two proposed components.
\subsubsection{Validation Results}
    As shown by Fig.~\ref{fig:case4}, both of MTOC-wLCA and MTOC-wGCF display higher multi-task performance than MTOC in two classification tasks. 
    Furthermore, they are less superior than DeepBroadcast, revealing the fact that these two components complement each other. 
    In addition, it is worth noting that the sub-encoder architecture is what these three systems have in common, evaluating the superiority of the architecture. 
    Overall, all the observations in four cases support the effectiveness of the proposed components and user-specific sub-encoding framework. 

\subsection{Ablation Study on Proposed Training Strategy}
\subsubsection{Setup}
    In the last case, based on case-III, we remove $\mathcal{L}_{overhead}$ in (\ref{broadcastIB_sketchy}) and replace the PFG and RP shown in Fig.~\ref{fig:pipeline} with a FC to construct DeepBroadcast-E2E. 
\subsubsection{Validation Results}
\begin{table}[htbp]
    \centering
    \caption{Average Accuracy in case-V}
      \begin{tabular}{|c|c|c|c|}
      \hline
      Networks & Task-1 & Task-2 & Task-3 \bigstrut\\
      \hline
      DeepBroadcast     & 94.67$\%$     & 86.44$\%$     & 72.43$\%$ \bigstrut[t]\\
      DeepBroadcast-E2E     & 93.29$\%$     & 83.64$\%$     & 65.27$\%$ \bigstrut[b]\\
      \hline
      \end{tabular}%
    \label{tab:Acc_case4}%
  \end{table}%
  
  As shown in Table.~\ref{tab:Acc_case4}, the system with the proposed training strategy, where the broadcast IB objective function is utilized, outperforms that with end-to-end training strategy in all three tasks. 
  This observation indicates that the proposed broadcast IB objective function supports fusion well with good feature compression to realize the full potential of DeepBroadcast. 

\section{Conclusion} \label{conclusion}
    In this paper, we proposed a multi-user broadcast semantic communication system named DeepBroadcast for simultaneously serving multiple tasks. 
    To balance task performance among multiple users, we decompose the channel encoding into two parts: compression and fusion to enlarge the degree of freedom of task-relevant information encoding. 
    In particular, we design task-channel-aware sub-encoders and channel-aware feature fusion sub-encoder to achieve compression and fusion, respectively. 
    Considering the adaptation of heterogeneous channels, we develop LCA and GCF to incorporate CSI into the semantic information space. 
    Moreover, besides introducing the information bottleneck theory, we locate the bottleneck in the proposed task-oriented broadcast system for better feature fusion. 
    Simulation results in 2$\&$3-user scenario have shown the outstanding multi-task performance of the proposed DeepBroadcast, especially in adverse channel conditions. 
    In particular, in Rician fading channel with intensity of -5dB, DeepBroadcast achieves $3.12\%$ higher accuracy than state-of-the-art system MTOC.

{
    \appendix[] 
	\appendices
	\section*{Proof of the Broadcast Information Bottleneck}
	It is critically important for a multi-user semantic communication system to achieve the highest multi-task performance with the least communication overhead over noisy channels meanwhile. 
    Nevertheless, such an optimization problem is contradictory.
    Achieving a high multi-task performance necessitates more symbols, which lead to increased communication overhead, to construct a complex feature space. 
    On the other hand, the transmitted signal with fewer symbols incurs lower communication overhead but also leads to worse multi-task performance due to the inadequate noise robustness.
    Fortunately, a potential equilibrium exists between the multi-task performance and communication overhead, which has been indicated by the IB theory \cite{alemi2016deep}.
    However, most of existing works such as \cite{shao2022task, shao2022taskvideo, shao2021learning} mainly focus on employing IB in single-user semantic communications.
    Hence, it is indeed important explore an approach to leverage advantages of IB in multi-user cases for high communication efficiency.

    Motivated by this, we extend the variational IB framework to adapt broadcast task-oriented semantic communication system. 
    In particular, a certain case is that only one user request to broadcast transmitter for communication. 
	Simply, we assume receiver-1 is the user.  
	For the transmitter and receiver-1, the broadcast communication in this moment can be simplified to that in one-to-one manner. 
	Therefore, the Markov Chain under this assumption can be given by
	\begin{equation}
        Y \to X \to Z \to \hat{Z} \to \hat{Y}. \label{markovchainO2O}
    \end{equation}
		
	According to the IB principle \cite{alemi2016deep}, the objective function of an one-to-one communication can be formulated as 
	\begin{equation}
		\mathcal L_{IB}(\bm{\phi}, \bm{\theta}) = -I(Z; Y)+\beta I(Z; X), \label{informationbottleneck}
	\end{equation}
	where $I(Z,Y)$ represents the mutual information between $Z$ and $Y$ and $I(Z,X)$ denotes the multual information between $Z$ and $X$. 
	Respectively, $I(Z,Y)$ is the objective function of maximizing the inference accuracy, which closely relates to the noise robustness, and $I(Z,X)$ is the objective function of minimizing the mutual information between the latent feature $Z$ and the source $X$. 
	Detailly, following the variational information bottleneck framework in \cite{alemi2016deep, shao2021learning}, we have 
	\begin{subequations}
		\label{distortion}
		\begin{align}
		I(Z;Y) &= \int p(\bm{y},\bm{z})\log{\frac{p(\bm{y}|\bm{z})}{p(\bm{y})}}d\bm{y}d\bm{z} \label{distortionA}\\
				&= \int p(\bm{y},\bm{z})\log{q_{\bm{\theta}}(\bm{y}|\bm{z})}d\bm{y}d\bm{z} \label{distortionB}\\ 
				&+ \int p(\bm{y},\bm{z})\log{\frac{p(\bm{y}|\bm{z})}{q_{\bm{\theta}}(\bm{y}|\bm{z})}}d\bm{y}d\bm{z} \label{distortionC}\\
				&+ H(Y) \label{distortionD}, 
		\end{align}
	\end{subequations}
	where $H(Y)$ is the entropy of $Y$ and $q_{\bm{\theta}}(\bm{y}|\bm{z})$ is a conditional distribution parameterized by $\bm{\theta}$. 
	As shown in (\ref{distortion}), $I(Z;Y)$ can be given by three monomial expressions including (\ref{distortionB}), (\ref{distortionC}), and (\ref{distortionD}). 
	(\ref{distortionB}) is a loglikelihood loss function and also called cross entropy and (\ref{distortionC}) is a Kullback-Leibler (KL)-divergence $D_{KL}(p(\bm{y}|\bm{z})||q_{\bm{\theta}}(\bm{y}|\bm{z}))$. 
	It is worth noting that $D_{KL}(p(\bm{y}|\bm{z})||q_{\bm{\theta}}(\bm{y}|\bm{z}))$ is always greater than or equal to zero and (\ref{distortionD}) is a constant given by practical data distribution.
	Generally, those optimization objectives determined by the practical data distribution are not optimizable. 
	Consequently, following \cite{alemi2016deep}, we have 
	\begin{equation}
		\label{accuracyKL}
		I(Z;Y) \geq \int p(\bm{y},\bm{z})\log{q_{\bm{\theta}}(\bm{y}|\bm{z})}d\bm{y}d\bm{z},
	\end{equation}
	showing that $I(Z;Y)$ can be optimized by maximizing a cross entropy function, which is the lower bound of $I(Z;Y)$. 
	According to Markov Property, we have 
	\begin{equation}
		p(\bm{y},\bm{z}) = \int p(\bm{x}, \bm{y}, \bm{z})d\bm{x} = \int p(\bm{x}, \bm{y})p(\bm{z}|\bm{x})d\bm{x},
	\end{equation}
	where $p(\bm{x},\bm{y})$ is determined by the source data distribution and it is constant.
	Then, leveraging the advantages of variation approaches, we map $\bm{x}$ to $\bm{z}$ parameterized by $\bm{\phi}$. 
	Furthermore, considering that semantic feature only contains discrete values, we have 
	\begin{subequations}
		\label{crossentropy}
		\begin{align}
			& \int p(\bm{y},\bm{z})\log{q_{\bm{\theta}}(\bm{y}|\bm{z})}d\bm{y}d\bm{z} \\
			&= \int p_{\bm{\phi}}(\bm{z}|\bm{x})p(\bm{x},\bm{y})\log{q_{\bm{\theta}}(\bm{y}|\bm{z})}d\bm{x}d\bm{y}d\bm{z} \label{CEA}\\
			&\simeq \int p_{\bm{\phi}}(\bm{z}|\bm{x})\log{q_{\bm{\theta}}\left(\bm{y}|\bm{z}\right)}d\bm{x}d\bm{y}d\bm{z} \label{CEB}\\
			&\simeq \mathbb{E}_{p_{\bm{\phi}(\bm{z}|\bm{x})}}\left(\log{q_{\bm{\theta}\left(\bm{y}|\bm{z}\right)}}\right), \label{CEC}
		\end{align}
	\end{subequations}
	indicating that we can improve the task performance by optimizing a cross entropy loss function (\ref{CEC}).  
	It is worth noting that each receiver is on duty of a unique task independently as we have claimed in Section~\ref{systemmodel}. 
	Consequently, it is feasible to improve the multi-task inference accuracy through optimizing $\mathcal{L}_{accuracy}$ in a broadcast semantic communication system.  
	
	We now focus on $I(Z;X)$ aiming at reducing the communication overhead. 
	Since this objective function can be simplified with a similar approach, we have 
	\begin{subequations}
		\label{rate}
		\begin{align}
		I(Z;X)  &= \int p(\bm{z},\bm{x})\log{\frac{p_{\bm{\phi}}(\bm{z}|\bm{x})}{q(\bm{z})}}d\bm{x}d\bm{z} \label{rateA}\\ 
				&+ \int p(\bm{z},\bm{x})\log{\frac{p(\bm{z})}{q(\bm{z})}}d\bm{x}d\bm{z}. \label{rateB}
		\end{align}
	\end{subequations}
	Thus, $I(Z;X)$ consists of two KL-divergence formulas, $D_{KL}(p_{\bm{\phi}}(\bm{z}|\bm{x})||q(\bm{z}))$ and $D_{KL}(p(\bm{z})||q(\bm{z}))$. 
	Notably, $D_{KL}(p(\bm{z})||q(\bm{z}))$ is also greater than or equal to zero. 
    Following \cite{shao2022task}, the variational marginal distribution $q(\bm{L}_{i})$ shoulde be treated as the centred isotropic Gaussian distribution $\mathcal N\left(\bm{L}_{i}|0, \bm{I}\right)$ and the KL-divergence can be simplified as 
    \begin{equation}
        \label{Dkl}
        \begin{aligned}
        & D_{KL}\left(p_{\bm{\phi}_i}(\bm{z}^r_{i}|\bm{x})\Big|\Big|q(\bm{z}^r_{i})\right) \\
        & =\sum_{k=1}^{c_2} \Big(\frac{\mu_{i,k}^{2}+\sigma_{i,k}^{2}-1}{2} - \log{\sigma_{i,k}}\Big).
        \end{aligned}
    \end{equation}
	In short, the system objective function includes the cross entropy part and the KL-divergence part with a Lagrange multiplier $\beta$.
}
\bibliography{twc_paper}    
\bibliographystyle{ieeetr}
\end{document}